\documentclass{article}

\usepackage{xcolor,colortbl}
\usepackage[x11names,dvipsnames]{xcolor}
\usepackage{graphicx}
\usepackage{amsmath}
\usepackage{amssymb}
\usepackage{booktabs}
\usepackage{amsthm}
\usepackage{comment}
\usepackage{multirow}
\usepackage{cancel}

\usepackage[margin=1in]{geometry}
\usepackage{natbib}
\setcitestyle{authoryear,round,citesep={;},aysep={,},yysep={;}}
\usepackage{hyperref}
\usepackage{wrapfig}
\usepackage{cleveref}
\usepackage{footnote}
\usepackage{pifont}

\makesavenoteenv{tabular}

\newcommand{\ul}[1]{{\underline{#1}}}
\newcommand{\bb}[1]{{\textbf{#1}}}

\newcommand{\y}{\boldsymbol{y}}

\newcommand{\w}{\boldsymbol{w}}
\newcommand{\bvar}{\boldsymbol{\varepsilon}}

\newcommand{\loss}[1]{\mathcal{L}_{\mathrm{#1}}}

\newcommand{\Id}{\mathbf{I}}
\newcommand{\const}{\mathrm{const}}
\newcommand{\x}{\boldsymbol{x}}
\DeclareMathOperator*{\argmin}{arg\,min}

\begin{document}

\begin{center}
    \section*{Learning to Recorrupt: Noise Distribution \\Agnostic Self-Supervised Image Denoising}
\end{center}

\begin{center}
    Brayan Monroy$^1$, Jorge Bacca$^{1,2}$, and Julian Tachella$^3$\\
    $^1$Universidad Industrial de Santander, Colombia \\
    $^2$GIPSA-Lab, Université Grenoble Alpes, Grenoble, France\\
    $^3$CNRS, ENS de Lyon, France \vspace{-0.5em}
\end{center}

\begin{center}
    \url{https://github.com/bemc22/Learning2Recorrupt}
\end{center}

\setlength{\parindent}{0pt}
\setlength{\parskip}{0.35em}

\begin{abstract}
Self-supervised image denoising methods have traditionally relied on architectural constraints, pseudo-pair constructions, or specialized loss functions to avoid the trivial identity mapping. Among these, approaches such as Noisier2Noise or R2R create training pairs by adding synthetic noise to noisy images. While effective, these recorruption-based approaches require precise knowledge of the noise distribution, which is often unavailable. We present Learning to Recorrupt (L2R), a self-supervised framework that does not require exact knowledge of the noise distribution. Our method introduces a learnable recorruptor jointly optimized with the denoiser through a min--max saddle-point objective. The proposed method achieves state-of-the-art performance among methods without prior knowledge of the noise distribution across unconventional and heavy-tailed noise distributions, such as log-gamma and Laplace, as well as spatially correlated noise, while obtaining competitive results on real-world denoising benchmarks characterized by signal-dependent noise.
\end{abstract}
\section{Introduction}
\label{sec:intro}

Image denoising remains an essential component of computer vision, serving as a critical preprocessing step in domains such as
computational photography~\citep{plotz2017benchmarking},
remote sensing~\citep{rasti2021image}, and medical imaging~\citep{kaur2023complete}. It operates as a fundamental block of contemporary image restoration algorithms~\citep{elad2023image, milanfar2025denoising} and generative models~\citep{ho2020denoising, li2025back}. While supervised image denoising has achieved remarkable success by exploiting large-scale datasets of clean--noisy pairs, its applicability is often restricted in real-world scenarios, where ground-truth images are either expensive or physically impossible to obtain~\citep{lehtinen2018noise2noise}. This limitation has driven the development of self-supervised denoising, which seeks to learn the underlying image manifold directly from noisy observations~\citep{zhang2023unleashing, diaz2026frozen}. The fundamental challenge in this paradigm is to define a training objective that avoids the trivial identity mapping (where the network simply learns to output the noisy input) while ensuring that the network achieves meaningful denoising.

Existing self-supervised strategies address this challenge through various structural or statistical priors. \emph{Noise2Noise}~\citep{lehtinen2018noise2noise} utilizes independent noisy images of the same scene, though such pairs are rarely available in practice. \emph{Blind-spot networks} (BSNs), such as those of \citet{laine2019high} and their variants~\citep{krull2019noise2void, wang2022blind2unblind,zhang2023mm, liu2025blind2sound}, impose architectural constraints that prevent a pixel from seeing itself, effectively leveraging spatial correlation to predict noise-free values. However, BSNs often suffer from reduced performance due to the missing central pixel information and are computationally expensive. Downsampling/masking-based strategies, such as Noise2Self~\citep{batson2019noise2self} and Neighbor2Neighbor (NBR2NBR)~\citep{huang2021neighbor2neighbor}, instead limit the information presented to the denoiser to construct self-supervised targets from a single noisy image, without imposing a blind-spot architecture. However, these approaches typically rely on a pixel-wise noise independence assumption, which limits their effectiveness in the presence of spatially correlated corruptions. Alternatively, methods based on \emph{Stein’s Unbiased Risk Estimator} (SURE) and its generalizations~\citep{hudson1978natural, eldar2008generalized} utilize the divergence of the denoiser to estimate the supervised loss~\citep{ramani2008monte}. While powerful, these divergence-based methods typically require explicit knowledge of the noise likelihood. \mbox{Recently, \emph{UNSURE}~\citep{tachellaunsure}} and other extensions~\citep{herbreteaudivergence} proposed a middle ground between BSNs and SURE by tackling the case where the noise distribution is known, but some of its parameters, such as the covariance, are unknown.

\begin{figure}[!t]
    \centering
    \includegraphics[height=0.31\linewidth]{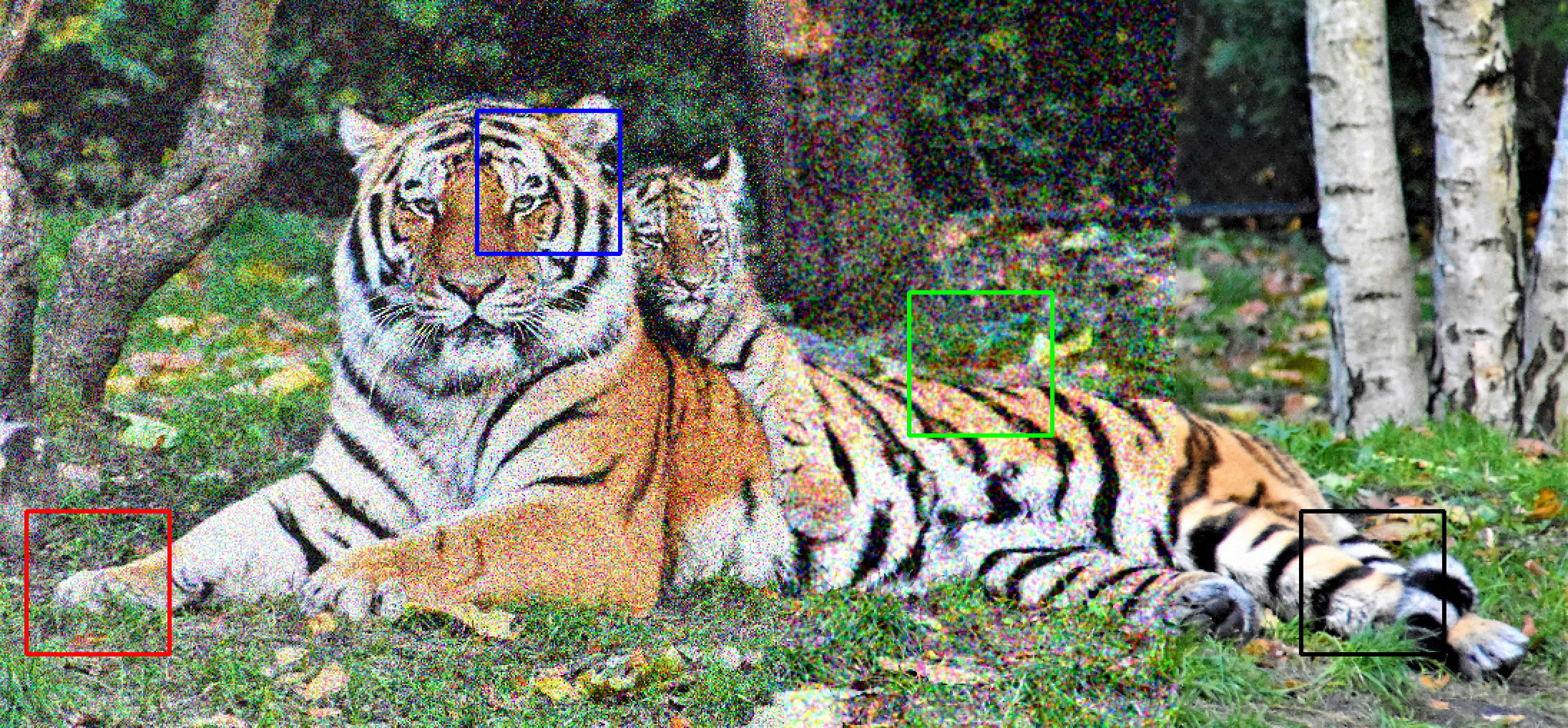}
    \includegraphics[height=0.31\linewidth]{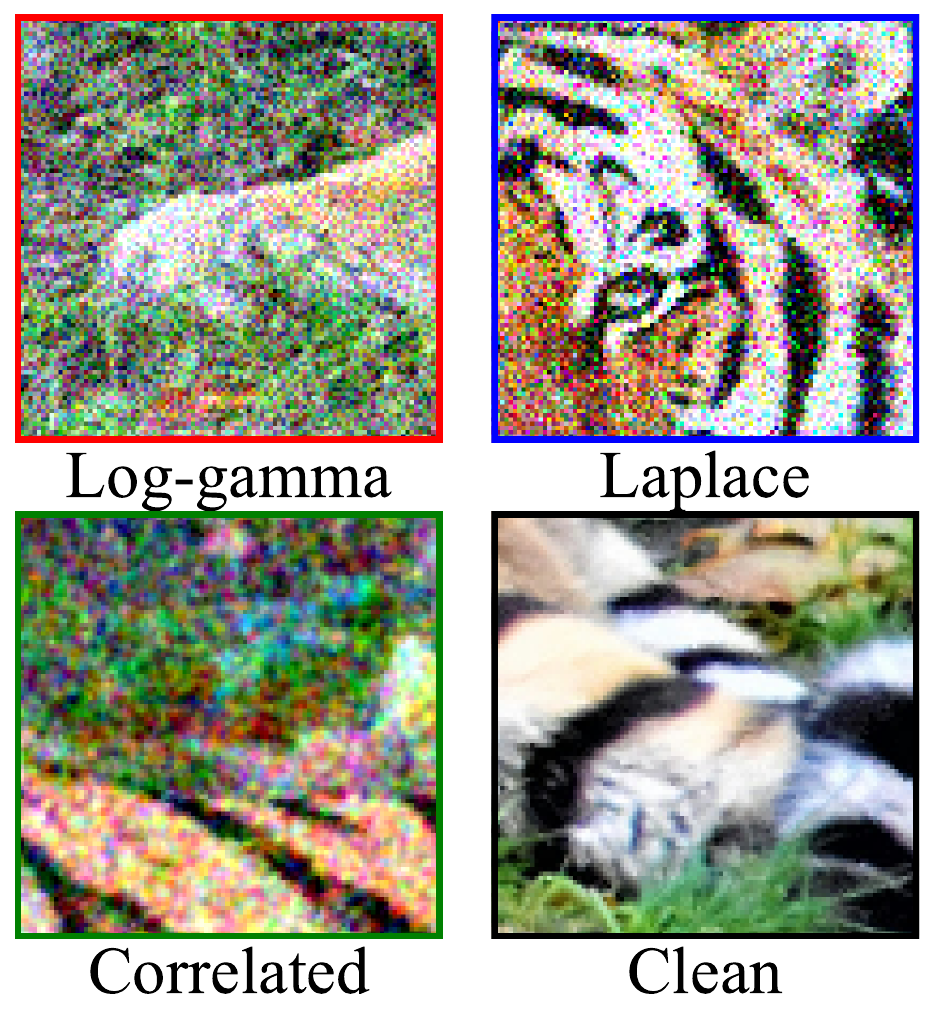} \vspace{-0.5em}
    \caption{\textbf{Non-Gaussian Image Denoising.} L2R is capable of handling complex non-Gaussian noise models without requiring prior knowledge of the noise distribution statistics, including log-gamma, Laplace, and correlated noise.}
    \label{fig:ngad}
\end{figure}

\begingroup\emergencystretch=3em
Recorruption-based methods, such as
Noisier2Noise~\citep{moran2020noisier2noise} and
R2R (Recorrupted-to-Recorrupted)~\citep{pang2021recorrupted},
construct training pairs by adding noise to a single noisy image. More recently, the
\emph{Generalized Recorrupted-to-Recorrupted framework} (GR2R)~\citep{monroy2025generalized} extended this mechanism
to additive and natural exponential-family noise, providing a
computationally efficient, architecture-agnostic, and unbiased proxy
for the supervised risk. However, GR2R requires knowledge of
the noise distribution to construct a valid recorruption; unknown
noise statistics therefore require manual tuning.
Related blind-denoising approaches, such as
GAN2GAN~\citep{cha2020gan2gan} and
DCD-Net~\citep{zou2023iterative}, learn noise models from single noisy
images and generate Noise2Noise pairs from progressively denoised estimates. Furthermore, complex noise can be modeled with normalizing
flows~\citep{abdelhamed2019noise}, while recent
noise-translation work learns to map real noise into Gaussian noise
for a pretrained denoiser~\citep{ha2026learning}; however, these
approaches typically require clean--noisy pairs, leaving a gap for
noise distribution agnostic approaches that work with single noisy
observations~\citep{lee2022ap}.\par\endgroup

To overcome this limitation, we propose \textbf{Learning to Recorrupt (L2R)}, a unified framework based on a noise distribution agnostic objective that extends recorruption-based self-supervision to settings with unknown noise statistics. L2R casts denoising as a min--max saddle-point problem in which the recorruption mechanism is treated as a learnable model whose parameterization can handle i.i.d., spatially correlated, and signal-dependent corruption structures. By employing a monotonic neural network, the recorruptor learns to map a standard Gaussian distribution into an auxiliary distribution satisfying the GR2R-compatible statistics. Rather than constraining the restoration architecture to enable self-supervision, L2R shifts the statistical constraint to the learned recorruptor, avoiding blind-spot or masking constraints on the denoiser.  This min--max optimization jointly adapts the auxiliary recorruption statistics and trains the denoiser to reduce the discrepancy between the self-supervised objective and its supervised counterpart, without requiring exact knowledge of the noise distribution. This enables robust denoising across noise types such as log-gamma, Laplace, and spatially correlated noise, as illustrated in Figure~\ref{fig:ngad}. The resulting framework retains the same objective across these corruption structures and the architectural neutrality of GR2R while providing a robust, data-driven approach for denoising across diverse and heavy-tailed noise regimes, with competitive performance on real-world signal-dependent noise.

\section{Related Work}
\enlargethispage{\baselineskip}

Given the rapid growth of self-supervised learning for denoising and imaging inverse problems, this section concentrates on the core methodological frameworks aligned with our formulation, namely supervised risk minimization, unbiased risk estimation, and recorruption-based training objectives. For a more comprehensive  overview (including additional self-supervised paradigms for noisy and incomplete measurements, their theoretical foundations, and practical imaging applications), we refer the reader to the recent survey~\citep{tachella2026self}.

\subsection{Supervised and Self-Supervised Learning}

The goal of supervised learning is to learn a deep denoiser operator $\hat{\x} = \hat{f}(\y)$ given clean--noisy paired data $(\y, \x)$. The population-optimal denoiser is defined by
\begin{equation}
   f^{*}=\argmin_f \; \mathbb{E}_{\x,\y} \, \loss{SUP}(\x,\y; f), \label{eqn:sup}
\end{equation} where $\loss{SUP}(\x,\y; f) = \Vert f(\y) - \x  \Vert_2^2$ is the MSE between the estimated and clean images\footnote{It is important to highlight that other cost functions can be used, such as the mean absolute error.}. Under the squared loss, the population optimum satisfies $f^{*}(\y)=\mathbb{E}\{\x\mid\y\}$. In practice, the population expectation is approximated with a finite dataset of $N$ samples, yielding the empirical estimator $\hat{f}$ through $\frac{1}{N}\sum_{i=1}^N \loss{SUP}(\x^{(i)},\y^{(i)};f)$. By simple algebraic manipulation, $\loss{SUP}$ can be split into self-supervised and supervised terms as follows 
\begin{equation} \label{eq: sup decomposition}
\loss{SUP}(\x,\y; f) = \Vert f(\y) -  \y  \Vert_2^2 + 2 f(\y)^\top( \y - \x) +  \const.
\end{equation} 
When there is no access to the set of clean images $\x$, self-supervised methods aim to approximate or eliminate the term that depends on $\x$, namely $f(\y)^\top(\y-\x)$, by building a self-supervised loss $\loss{SELF}$ such that \begin{equation}
    \mathbb{E}_{\y} \loss{SELF}(\y; f) = \mathbb{E}_{\y,\x}\loss{SUP}(\x,\y; f) +  \const. 
\end{equation}

\subsection{Stein's Unbiased Risk Estimator and Extensions}

A more realistic scenario is the self-supervised case, with access only to unpaired noisy measurements. Assuming Gaussian observations $\y|\x \sim \mathcal{N}(\x,\sigma^2 \mathbf{I})$ and a weakly differentiable estimator $f:\mathbb{R}^n\!\to\!\mathbb{R}^n$, Stein's unbiased risk estimator (SURE) defines the self-supervised loss as \begin{equation}
\loss{SURE}(\y;f)
=
\Vert f(\y) - \y \Vert_2^2
+
2\sigma^2\,\mathrm{div}\,f(\y),
\end{equation}
with $\mathrm{div}\,f(\y):=\sum_{i=1}^{n} \frac{\partial f_i}{\partial y_i}(\y)$  which relies on Stein's identity~\citep{stein1972bound} \begin{equation}
\mathbb{E}_{\y|\x}\!\left[\big(f(\y)-\x\big)^\top(\y-\x)\right]
=
\sigma^2\,\mathbb{E}_{\y|\x}\!\left[\mathrm{div}\,f(\y)\right],
\label{eq:stein_identity_sure}
\end{equation}
so that $\mathbb{E}_{\y|\x}[\loss{SURE}(\y;f)]$ is an unbiased estimator of the supervised risk $ \mathbb{E}_{\y|\x} \mathcal{L}_{\text{SUP}}$ up to $f$-independent constants, with the minimizer approaching the MMSE estimator.
Noteworthy extensions beyond i.i.d.\ Gaussian noise include Hudson's lemma~\citep{hudson1978natural} and GSURE~\citep{eldar2008generalized}, which extends SURE to the exponential family.

For correlated Gaussian noise $\y|\x\sim\mathcal{N}(\x,\mathbf{\Sigma})$, the corresponding Stein identity replaces
$\sigma^2\,\mathrm{div}\,f(\y)$ with a $\mathbf{\Sigma}$-weighted divergence term $\mathrm{tr}\!\left(\mathbf{\Sigma}\,\frac{\partial f}{\partial \y}(\y)\right)$.

\textbf{UNSURE}~\citep{tachellaunsure} further addresses \emph{unknown} noise statistics by replacing the explicit variance-dependent correction with a constraint on the expected divergence.
In its basic form, UNSURE considers the denoising objective
\begin{equation} 
\min_{f}\;
\mathbb{E}_{\y}\,\|f(\y)-\y\|_2^2
\quad\text{s.t.}\quad
\mathbb{E}_{\y}\big[\mathrm{div}\,f(\y)\big]=0,
\label{eq:unsure_zed_constraint}
\end{equation}
referred to as the \emph{zero expected divergence} constraint.
Equivalently, a Lagrangian (saddle) form is
\begin{equation}
\min_{f}\;\max_{\eta\in\mathbb{R}}\;
\mathbb{E}_{\y}\,\|f(\y)-\y\|_2^2
+2\eta\,\mathbb{E}_{\y}\big[\mathrm{div}\,f(\y)\big].
\label{eq:unsure_lagrangian}
\end{equation}

 \subsection{Generalized Recorrupted-to-Recorrupted}

The GR2R framework extends the R2R approach~\citep{pang2021recorrupted}, originally proposed for Gaussian noise, to additive noise and the natural exponential family (including Poisson, Gamma, hypergeometric, among others), exhibiting a close connection with SURE. In the additive noise setting, given a noisy observation $\y=\x+\boldsymbol{\varepsilon}$, where $\boldsymbol{\varepsilon}$ denotes the measurement noise, GR2R generates a recorrupted pair as follows
\begin{equation} \label{r2req}
    \y_1 = \y + \tau \boldsymbol{\omega}, \quad \y_2 = \y - \boldsymbol{\omega} / \tau,
\end{equation}
where $\tau>0$ and $\boldsymbol{\omega}$ is drawn independently of $\boldsymbol{\varepsilon}$. In the original R2R setting (Gaussian noise with known variance), one may take $\boldsymbol{\omega} \sim \mathcal{N}(\mathbf{0}, \sigma^2 \Id)$. In the general additive non-Gaussian case, $\boldsymbol{\omega}$ is instead chosen so that its low-order moments match those of $\boldsymbol{\varepsilon}$, e.g.,

\begin{equation} \label{eq:moment_match_general} \mathbb{E}_{\varepsilon_i,\omega_i} \left[ \left( \varepsilon_i-\frac{\omega_i}{\tau} \right) \left( \varepsilon_i+\tau\omega_i \right)^k \right] = 0, \qquad i=1,\ldots,n, \quad k=0,\ldots,K . \end{equation}

Here, $K$ denotes the highest moment order enforced by the matching conditions.

Then, the self-supervised denoising loss is defined as \begin{equation}\label{eq:r2r_loss}
\begin{aligned}
    \loss{GR2R} (\y;f) &= \mathbb{E}_{\y_1,\y_2|\y} \big\|  f(\y_1) - \y_2 \big\|_2^2. 
\end{aligned}
\end{equation} Under moment-matching conditions such as those in Eq.~\eqref{eq:moment_match_general} (for instance, in the case of Gaussian noise with variance-matching $\boldsymbol{\omega}\stackrel{d}{=}\boldsymbol{\varepsilon}$), the population GR2R objective becomes, up to an additive constant, equivalent to a supervised denoising objective on $\y_1$, that is, $\mathbb{E}_{\y}\loss{GR2R}(\y;f) \propto \mathbb{E}_{\y_1|\x}\mathcal{L}_{\text{SUP}}(\x,\y_1;f)$, with the denoiser model satisfying $f^*(\y_1)\approx\mathbb{E}\{\x\mid \y_1\}$.

\section{Learning to Recorrupt}

We consider the additive denoising model
\begin{equation}
    \y = \x + \bvar,
    \qquad \bvar \perp \x,
    \qquad \mathbb{E}[\bvar]=0,
    \label{eq:noise_model}
\end{equation}
where $\y\in\mathbb{R}^n$ denotes the noisy observation, $\x\in\mathbb{R}^n$ is the unknown clean image, and $\bvar\sim p_{\bvar}$ follows an \emph{unknown} distribution. Specifically, let $\w\sim\mathcal{N}(\mathbf{0},\mathbf{I})$ and define
\begin{equation}
    \bvar = g(\w),
\end{equation}
for a map $g:\mathbb{R}^n \rightarrow \mathbb{R}^n$ that pushes $\mathcal{N}(\mathbf{0},\mathbf{I})$ forward to the target noise distribution, i.e., $g_{\#}\mathcal{N}(\mathbf{0},\mathbf{I})=p_{\bvar}$, where $g_{\#}$ denotes the pushforward measure induced by $g$. With this reparameterization, the observation model becomes
\begin{equation}
    \y = \x + g(\w), 
    \qquad \w\sim\mathcal{N}(\mathbf{0},\mathbf{I}),
\end{equation}
so the unknown corruption is represented as the pushforward of a Gaussian distribution. We follow the intuition of GR2R and construct a recorrupted image $\y_1$ from the single noisy image $\y$. In classical GR2R, the choice of recorruption relies on explicit knowledge of the noise distribution. 
To overcome this limitation, we propose a learned alternative by introducing an auxiliary recorruption mapping $h:\mathbb{R}^n \rightarrow \mathbb{R}^n$ constrained to a class $\mathcal{H}$ of admissible vector-valued recorruption maps that are (a.e.) differentiable, and whose outputs have zero mean, i.e., $\mathbb{E}_{\w'\sim\mathcal{N}(\mathbf{0},\mathbf{I})}[h(\w')]=\mathbf{0}$. This mapping generates a recorrupted observation as
\begin{equation} \label{eq:l2r_recorruption}
    \y_1 = \y + \tau h(\w'),
\end{equation}
where $\w' \sim\mathcal{N}(\mathbf{0},\mathbf{I})$ and $\tau>0$ controls the strength of the recorruption.

\begin{figure}[t]
    \centering
    \includegraphics[width=\linewidth]{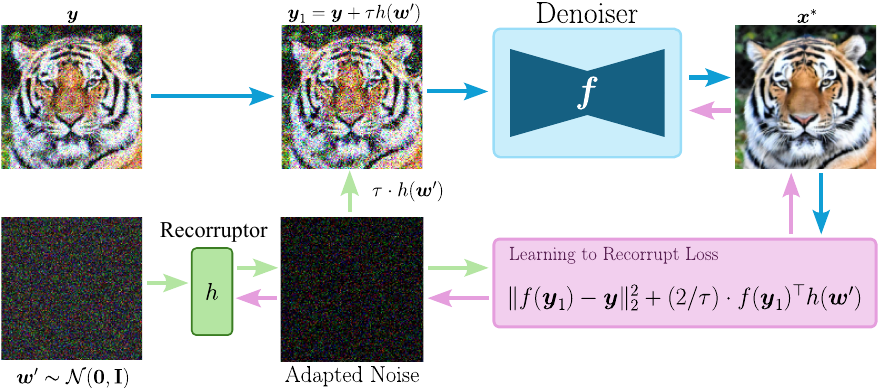} \vspace{-1.5em}
    \caption{\textbf{Learning to Recorrupt.} Given a noisy image $\y$ with unknown noise, L2R generates a recorrupted version $\y_1$ through a learned recorruptor $h$ to enable self-supervised denoiser training.}
    \label{fig:AR2R}
\end{figure}

As in Eq.~\eqref{eq: sup decomposition}, we can expand the simple measurement consistency loss into a supervised term and a term measuring correlation with the noise:
\begin{equation}
\begin{aligned}
   &\mathbb{E}_{\y_1,\y} \|f(\y_1) - \y\|^2_2  \\
   &= \mathbb{E}_{\y_1,\x}\|f(\y_1) - \x\|^2_2 - 2 \, \mathbb{E}_{\x,\w,\w'} f(\x+ g(\w)  + \tau h(\w'))^{\top} g(\w) + \text{const.}
\end{aligned}
\end{equation}

Since $g$ is unknown, we seek to control this correlation through a family of constraints:
\begin{equation} \label{eq: constrained minim}
   f^{*} = \arg\min_f \; \mathbb{E}_{\y_1,\y}\|f(\y_1) - \y\|^2_2 \quad \text{s.t.} \; \mathbb{E}_{\y,\w'}f(\y+\tau h(\w'))^{\top} h(\w') = 0 \quad \forall h \in \mathcal{H}. 
\end{equation}

An auxiliary recorruption map is compatible when its correlation term is proportional to the unknown correlation with $g(\w)$, according to Eq.~\eqref{eq:moment_match_general}. If $\mathcal{H}$ contains such a map $h_c$, then $\mathbb{E}_{\x,\w,\w'} f^{*}(\x+g(\w)+\tau h_c(\w'))^{\top}g(\w)=0$, and, for this fixed compatible map, Eq.~\eqref{eq: constrained minim} is equivalent to the \emph{supervised} constrained minimization problem:

\begin{equation} \label{eq: sup constrained minim}
    f^{*} = \arg\min_f \; \mathbb{E}_{\y_1,\x}\|f(\y_1) - \x\|^2_2 \quad \text{s.t.} \; \mathbb{E}_{\y,\w'}f(\y+\tau h(\w'))^{\top} h(\w') = 0 \quad \forall h \in \mathcal{H}.
\end{equation}

Further analysis is provided in Appendix~\ref{sec:sm_eq16_eq17}.
Assuming that the output of $f$ is bounded and that $\mathcal{H}$ is a conic set\footnote{For all $\alpha>0$ and $h\in\mathcal{H}$, the scaled map $\alpha h$ also belongs to $\mathcal{H}$. For the additive parameterization in Eq.~\eqref{eq:recorruptor_additive}, this follows by scaling the learnable kernel $\boldsymbol{k}_h$ by $\alpha$.}, the constraints in Eq.~\eqref{eq: constrained minim} motivate the following adversarial saddle-point objective:
\begin{equation} \label{eq: lagrange minim}
\begin{gathered}
    \min_f\max_{h\in \mathcal{H}}
    \; \mathbb{E}_{\y}
    [\mathcal{L}_{\text{L2R}}(\y;f,h)], \\
    \mathcal{L}_{\text{L2R}}(\y;f,h)
    :=
    {
    \mathbb{E}_{\w'}
    \left[
        \|f(\y+\tau h(\w'))-\y\|_2^2
        +
        \frac{2}{\tau}
        f(\y+\tau h(\w'))^\top h(\w')
    \right]}.
\end{gathered}
\end{equation}

In practice, we represent both $f$ and $h$ as neural networks and alternate between gradient descent with respect to $f$ and gradient ascent with respect to $h$, as in adversarial learning~\citep{arjovsky2017towards}. We use a stop-gradient surrogate for the recorruptor update. Unlike blind-spot networks~\citep{laine2019high}, which impose architectural constraints to prevent access to the central pixel, the denoiser $f$ in L2R can be parametrized using any architecture.

\textbf{Designing the learnable recorruptor.}
The architecture and capacity of the recorruptor $h$ determine the range of auxiliary noise statistics it can represent, providing a way to control the trade-off between robustness to unknown noise distributions, and the gap with respect to the optimal supervised loss~\citep{tachellaunsure,tachella2026self}. Therefore, we propose four $h$ variants for two scenarios encountered in practice: {non-signal-dependent noise}, with (1.a) an elementwise recorruptor and (1.b) a spatial recorruptor; and {signal-dependent noise}, with (2.a) a local recorruptor and (2.b) a global recorruptor.

For the \textbf{non-signal-dependent scenario}, we define
\begin{equation}
	h(\w')
	=
	\boldsymbol{k}_h * \mathrm{N}\!\left(\mathrm{mMLP}(\w')\right),
	\label{eq:recorruptor_additive}
\end{equation}
where $\mathrm{mMLP}$ is a monotonic network~\citep{runje2023constrained}, inspired by the inverse-transform principle that a monotonic map can transform a Gaussian variable into a target distribution. The operator $\mathrm{N}(\cdot)$ normalizes its output to zero mean and unit variance. Variant (1.a) uses a $1\times1$ kernel $\boldsymbol{k}_h$ to model non-correlated noise, while variant (1.b) uses a spatial kernel to model spatially correlated noise.

For the \textbf{signal-dependent scenario}, we define
\begin{equation}
	h(\w',\y)
	=
	s(\y)\odot \tilde h_{\y} (\w'),
	\label{eqn:signaldepent}
\end{equation}
where $s(\y)$ sets the recorruption scale as a function of the observation, and $\tilde h_{\y}$ follows the architecture in Eq.~\ref{eq:recorruptor_additive}. In the local variant (2.a), the weights of $\tilde h_{\y}$ are shared across images; in the global variant (2.b), a hypernetwork predicts image-specific weights from $\y$. Further implementation details are provided in Appendix~\ref{sec:noisemodels}.

\textbf{Interpretation of $h$ at convergence.}
At a saddle point $(f^*,h^*)$, we can rewrite Eq.~\eqref{eq: lagrange minim} as
\begin{align*}
    \min_f \mathbb{E}_{\y_1,\x}\|f(\y_1) - \x\|_2^2
    + 2\,\mathbb{E}_{\x,\w,\w'}
    f(\y+\tau h^{*}(\w'))^{\top}
    \left(
        \frac{1}{\tau}h^{*}(\w') - g(\w)
    \right).
\end{align*}
The second term provides a measure of alignment with the supervised objective, which we examine empirically through the following relation:
\begin{equation}
\label{eqn:equilibrium}
    \mathbb{E}_{\x,\w,\w'}
    f^{*}(\y+\tau h^{*}(\w'))^{\top}
    \frac{1}{\tau}h^{*}(\w')
    \approx
    \mathbb{E}_{\x,\w,\w'}
    f^{*}(\y+\tau h^{*}(\w'))^{\top}
    g(\w).
\end{equation}
Thus, the learned auxiliary corruption tracks the true-noise contribution through the denoiser-weighted correlation. This relation is evaluated numerically in Appendix~\ref{sec:moment}.

\textbf{Relationship to GR2R and UNSURE.}
The proposed self-supervised objective can be seen as a generalization of the UNSURE loss in Eq.~\eqref{eq:unsure_zed_constraint}. L2R uses a more flexible set of constraints parametrized by a (monotonic) neural network. The method can also be linked to GR2R in Eq.~\eqref{eq:r2r_loss}, as the loss in Eq.~\eqref{eq: lagrange minim} can be rewritten as 
 \begin{align*}
    \min_f \max_h \; \mathbb{E}_{\y_1,\y_2}\|f(\y_1) - \y_2\|^2_2 
    + \mathbb{E}_{\y, \w'}(\y + \y_2)^\top h(\w') / \tau,
 \end{align*}
 with $\y_2 = \y - \frac{1}{\tau}h(\w')$, where the recorruption process is learned instead of being fixed. In GR2R, the second term is omitted because it is not optimized. See Appendices~\ref{sec:gr2r_relation} and~\ref{sec:unsure_relation} for details.

\section{Simulations and Results}
\label{sec:results}

This section evaluates the proposed \textbf{L2R} framework in the distribution-agnostic self-supervised denoising setting, where the corruption follows an unknown noise distribution and clean targets are unavailable. We compare it with noise distribution-aware recorruption methods \textbf{GR2R}~\citep{monroy2025generalized} and \textbf{R2R}~\citep{pang2021recorrupted}, divergence-based estimators \textbf{(PG)-SURE}~\citep{ramani2008monte, le2014unbiased} and \textbf{(PG)-UNSURE}~\citep{tachellaunsure}, noise distribution agnostic baselines \textbf{NBR2NBR}~\citep{huang2021neighbor2neighbor} and \textbf{Blind2Sound}~\citep{liu2025blind2sound}, and \textbf{Supervised} training as an upper bound. Note that GR2R can be seen as an oracle in our framework: our method reduces to GR2R when the learned recorruption operator matches the true noise-induced recorruption assumed in GR2R. 

\textbf{Datasets.}
Experiments are conducted on BSDS500 following the standard $200/100/200$ train/validation/test split, and all quantitative results are reported on the test subset. During training, images are randomly cropped into patches of spatial resolution $256\times256$ pixels.  To assess robustness under distribution shift, the DIV2K dataset is employed exclusively for evaluation: performance is measured on the 100 validation images, using $512\times512$ pixel center crops. 

\textbf{Training setup.}
For synthetic-noise experiments, all methods employ a
lightweight DRUNet~\citep{zhang2021plug} variant, comprising
approximately $2$M trainable parameters.
{A single noisy realization is generated for each
training image and kept fixed during training.}
The models are trained with a batch size of 32 for 2000 epochs using
the AdamW optimizer with a weight decay of $0.01$ and a cosine-annealed
learning-rate schedule decreasing from $10^{-4}$ to $10^{-6}$, with
$\tau=0.5$.
{For experiments on SIDD~\citep{abdelhamed2018sidd}, we adopt the U-Net
architecture and training schedule of Blind2Sound~\citep{liu2025blind2sound}, using Adam for 400
epochs with $\tau=1.0$ and an initial learning rate of $10^{-4}$,
halved every 80 epochs. The recorruptor learning rate is set to half
that of the denoiser. To reduce training cost, stop-gradient is applied to the input of the denoiser $f(\mathrm{sg}[\y_1])$, preventing gradients
through the denoiser with respect to $\y_1$ while keeping a similar performance, as shown in Appendix~\ref{sec:optimsetup}. The implementation is based on the \texttt{DeepInverse}
library~\citep{tachella2025deepinverse}.}

\textbf{Noise settings.} We evaluate L2R under zero-mean log-gamma and Laplace noise,
spatially correlated Gaussian noise, and signal-dependent
Poisson--Gaussian noise. Across these settings, the L2R objective
remains unchanged, while the recorruptor structure encodes only broad
assumptions such as element-wise, spatially correlated, or
signal-dependent corruption. The noise parameters and distribution
statistics are not provided to L2R. Exact generation procedures and
recorruptor settings are given in Appendix~\ref{sec:noisemodels}.
Additional visualizations of the learned recorruption maps and
spatial kernel are provided in Appendix~\ref{sec:implicit_moments_kernel}.

\begin{table*}[!t] \centering 
\caption{\bb{Non-Gaussian Denoising.} Performance under different types of non-Gaussian noise. Best and second-best self-supervised results are shown in bold and underlined, respectively. Gray rows are reference methods excluded from the ranking. \label{tab:nongaussian}}  
\resizebox{\linewidth}{!}{\begin{tabular}{lc|ccc|ccc|cccc} 
\toprule
 \multirow{2}{*}{\bb{Method}}
 &  $p_{\bvar}$
 & \multicolumn{3}{c}{\bb{Log-gamma Noise}} 
 & \multicolumn{3}{c}{\bb{Laplace Noise}}
 & \multicolumn{3}{c}{\bb{Correlated Noise}} \\ 
 & known? &  $\ell$    & \bb{BSDS500} & \bb{DIV2K} 
 & $b$          & \bb{BSDS500} & \bb{DIV2K}
 & $\sigma$     & \bb{BSDS500} & \bb{DIV2K} 
 \\ \midrule \midrule 
\rowcolor{gray!15} 
  Supervised  
  &  no    &   &  31.44/0.898 & 31.96/0.894 
  &  &  29.48/0.856 & 30.21/0.859
  &  &   30.56/0.897   & 30.75/0.887  
  \\    \rowcolor{gray!15} 
  GR2R  
  &   yes  &  &  31.14/0.889  & 31.98/0.893
  &   & 28.10/0.803 & 28.74/0.807 
  &   & 30.43/0.893  & 30.61/0.884 
  \\   \rowcolor{gray!15} 
 SURE
  &   yes  &  &  28.33/0.796  & 28.61/0.781  
  &   & 27.07/0.748 & 27.60/0.747 
  &   &  30.37/0.891 & 30.60/0.882  
  \\     \rowcolor{gray!15} 
  R2R 
  & yes & &  28.63/0.809 & 29.03/0.799   
  &  &  27.85/0.797 & 28.52/0.804  
  &   &  24.52/0.653   & 24.97/0.634 
  \\   UNSURE
  &  no   & 1.0   &  27.83/0.756 & 28.11/0.737 
  &  0.1  &   22.93/0.615 &  23.42/0.639
  &  0.2  &   \ul{29.56}/\ul{0.841} & \ul{29.56}/\ul{0.821} 
  \\   NBR2NBR  
 & no &    &  29.45/0.830   & 29.78/0.818
 &  & 27.39/0.746 &  27.86/0.740
 &   &  25.16/0.680 & 25.53/0.657
 \\  Blind2Sound
 & no &    &  \ul{29.61}/\ul{0.857}   &  \ul{30.11}/\ul{0.849}
  &  &  \ul{27.86}/\ul{0.802}    &  \ul{28.43}/\ul{0.801} 
 &  &   24.20/0.643    &  24.53/0.620 
 \\   L2R (Ours) 
  & no &    &  \textbf{30.77/0.871}  &  \textbf{31.23/0.865}   
  & &  \textbf{28.40/0.823}  & \textbf{29.12/0.830}
  &   &  \textbf{29.85/0.863} & \textbf{29.75/0.836}
  \\      \midrule    \rowcolor{gray!15} 
    Supervised  
   &  no  &    & 33.40/0.934 & 33.57/0.926 
   &  &  32.46/0.915  & 32.84/0.909   
    &  &  34.48/0.949  & 34.25/0.938 
   \\  \rowcolor{gray!15} 
GR2R    
  & yes  &   &  32.34/0.915   & 33.01/0.915  
  &  &  31.90/0.901 & 32.29/0.896
   &  &  34.16/0.945   & 34.03/0.936 
   \\  \rowcolor{gray!15} 
SURE    
 & yes  &   &  28.26/0.804  & 28.73/0.792  
  &  &  31.18/0.879 & 31.42/0.869
  &   &  34.41/0.948 & 34.20/0.937 
  \\    \rowcolor{gray!15} 
  R2R    
  & yes   &  &  27.93/0.781  & 28.33/0.768  
  &  &     31.30/0.885 & 31.61/0.878  
  &   &  29.64/0.853  & 30.00/0.837 
  \\  UNSURE
 &  no  &  0.1    &  26.95/0.745  & 27.22/0.719   
 &  0.05   &  30.10/0.843  & 30.27/0.827 
 &  0.1    &  \ul{33.16}/\ul{0.915}  & \ul{32.58}/\ul{0.891}
 \\  NBR2NBR 
 &  no &   &  \ul{30.63}/0.864  & \ul{30.90}/0.850 
 &  &   \ul{31.49}/\bb{0.888} & \bb{32.10}/\bb{0.886}
 &   &  30.58/0.863 & 30.57/0.836
  \\  Blind2Sound 
 & no &    &  30.39/\ul{0.878}   &  30.81/\ul{0.867} 
 &  &  30.89/0.878   & 31.44/\ul{0.879}
 &  &  28.33/0.781   &  28.70/0.756 
 \\   L2R (Ours)
 &  no &   &  \textbf{32.29/0.901} & \textbf{32.40/0.887}
 &   &  \bb{31.59}/\ul{0.880}  & \ul{31.79}/0.868
 &   &  \bb{33.31/0.921} & \bb{32.87/0.900} 
 \\   \bottomrule
\end{tabular} }
\end{table*}

\subsection*{Non-Gaussian Denoising.}
\label{sec:results_nongaussian}

\begin{figure*}[t]
    \centering
    \includegraphics[width=\linewidth]{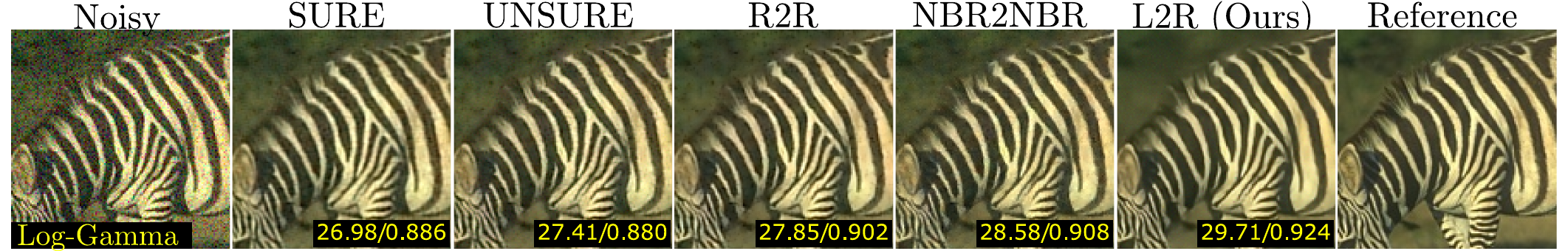}
    \includegraphics[width=\linewidth]{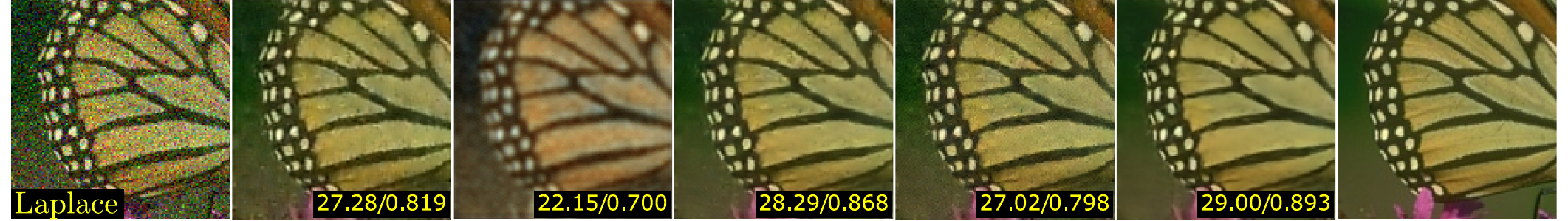}
    \includegraphics[width=\linewidth]{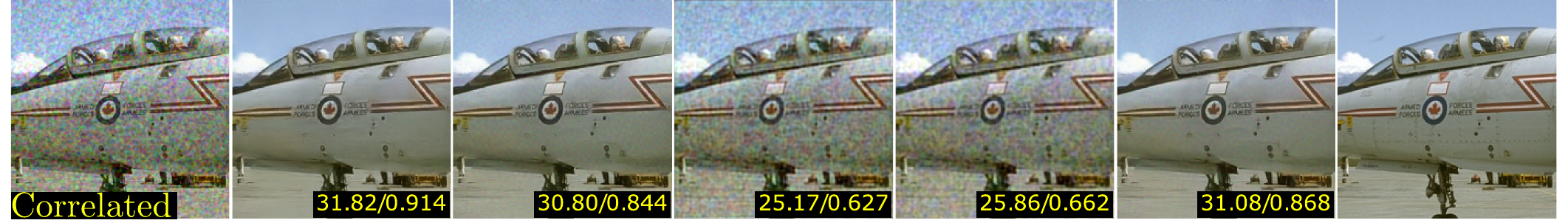} \vspace{-2em}
    \caption{\textbf{Non-Gaussian Denoising.} Visual comparison on representative BSDS500 crops corrupted by log-gamma, Laplace, and spatially correlated noise. Columns show the noisy input, SURE, UNSURE, R2R, NBR2NBR, and the proposed L2R, followed by the ground truth. Numbers in each panel report PSNR (dB) / SSIM. }
    \label{fig:visualnongaussian} \vspace{-0.5em}
\end{figure*}

Table~\ref{tab:nongaussian} reports quantitative results on three representative non-Gaussian noise types, log-gamma (heavy-tailed), Laplace, and spatially correlated noise, evaluated on BSDS500 and DIV2K.  In the case of log-gamma noise, L2R attains the highest PSNR among distribution-agnostic methods, reaching 30.77~dB (BSDS500) / 31.23~dB (DIV2K) in the first setting ($\ell=1.0$) and improving to 32.29~dB / 32.40~dB in the second ($\ell=0.1$). The strongest noise distribution-aware competitor (GR2R) remains slightly higher in PSNR and better in SSIM than L2R across both datasets, while UNSURE, NBR2NBR, and Blind2Sound exhibit larger gaps, particularly under the heavier-tailed regime. Overall, these results indicate that L2R preserves strong performance under heavy-tailed corruptions without requiring access to $p_{\bvar}$.

Regarding Laplace noise, L2R also provides the best PSNR in the first setting, achieving 28.40~dB on BSDS500 and 29.12~dB on DIV2K. In the second setting, L2R remains among the top-performing methods, reaching 31.59~dB and 31.79~dB, with only small gaps to the best model-aware alternatives. Notably, the second-best method changes across Laplace settings (and across datasets), whereas L2R remains among the strongest distribution-agnostic methods across these settings.

For spatially correlated noise, SURE yields the highest PSNR when the correlation kernel is known, but L2R is the best self-supervised alternative without access to $p_{\bvar}$. In the first setting, L2R reaches 29.85~dB (BSDS500) and 29.75~dB (DIV2K), and in the second it improves to 33.31~dB and 32.87~dB, substantially outperforming other distribution-agnostic baselines and narrowing the gap to SURE. Overall, while the strongest competing baseline varies with the noise distribution, L2R remains the most competitive self-supervised option across all non-Gaussian regimes considered.

The visual comparison in Figure~\ref{fig:visualnongaussian} supports the quantitative trends. Under log-gamma and Laplace noise, L2R better suppresses granular artifacts while preserving high-frequency textures (e.g., zebra stripes and fine edges) compared to SURE/UNSURE and prior self-supervised baselines. Under correlated noise, L2R reduces spatially structured residuals more effectively than distribution-agnostic competitors, producing reconstructions that are visually closer to the ground truth while avoiding oversmoothing. Overall, L2R provides the best robustness--performance trade-off: it achieves state-of-the-art results among methods that do not assume access to the underlying noise distribution while remaining competitive with noise distribution-aware approaches in structured noise.

\subsection*{Real-World Denoising.}

\begin{figure}[!t]
    \centering
    \includegraphics[width=\linewidth]{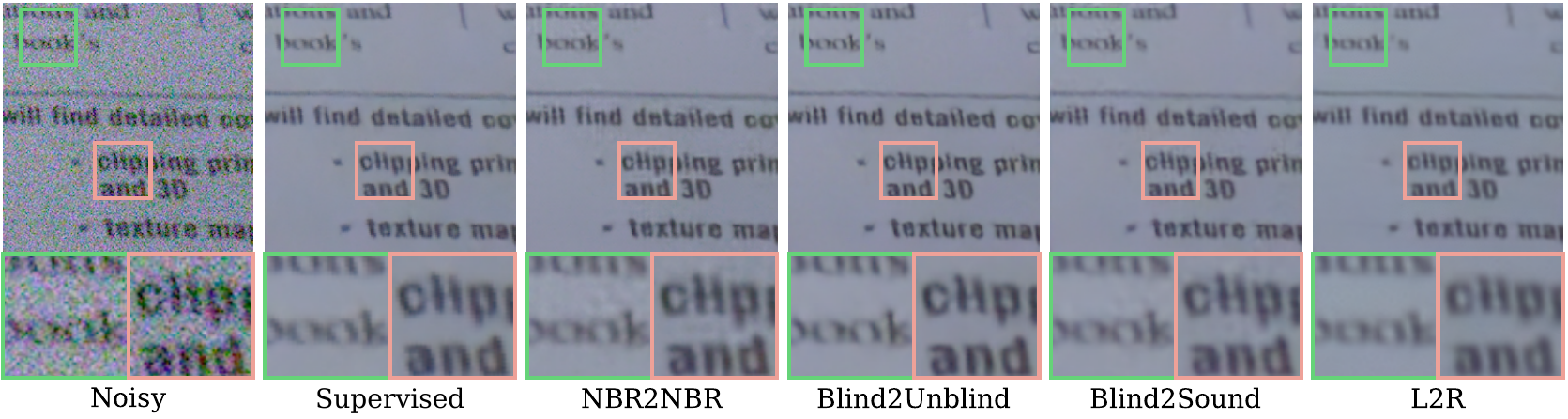}\vspace{-1em}
    \caption{Visualization of qualitative results on a sample image from the SIDD benchmark.}
    \label{fig:realsidd}
\end{figure}

Table~\ref{tab:sidd} reports results on SIDD, which contains real, spatially varying, and signal-dependent camera noise. The estimated training cost is computed as the ratio of the mean training image throughput of Supervised to that of each method, with Supervised normalized to $\times1$. L2R with the global recorruptor from Eq.~\eqref{eqn:signaldepent} achieves 51.35 dB and 0.992 SSIM on the validation set and 50.60 dB and 0.991 SSIM on the benchmark. Figure~\ref{fig:realsidd} supports these quantitative results: L2R removes strong chromatic noise while preserving text contours, producing a reconstruction visually comparable to Supervised and specialized blind-spot methods. L2R outperforms NBR2NBR by 0.29 dB on validation and 0.13 dB on the benchmark, while reducing the estimated training cost from $\times2$ to $\times1.4$. It remains within 0.15 dB of Blind2Sound on validation and 0.32 dB on the benchmark at approximately one-third of its estimated training cost. Compared with Blind2Unblind, L2R obtains similar accuracy with approximately $12\times$ lower training cost. Both L2R and Blind2Unblind attain the same benchmark SSIM at the reported precision.

Notably, L2R closely matches both the Supervised and Noise2Noise reference objectives. On the SIDD benchmark, L2R reaches 50.60 dB, compared with 50.61 dB for Supervised and 50.62 dB for Noise2Noise, while all three obtain an SSIM of 0.991. On validation, L2R slightly exceeds both references in PSNR and SSIM. Since Supervised and Noise2Noise share the same population MSE minimizer under standard zero-mean noise assumptions, these results indicate that the learned recorruption objective can recover performance comparable to clean-target and paired-noisy training without either form of supervision or a blind-spot architecture.

\begin{table}[!htbp]
    \centering \footnotesize
    \caption{Real-world denoising performance on SIDD. Training cost is reported relative to Supervised training. Best and second-best self-supervised results are highlighted.\label{tab:sidd}}
    \resizebox{0.6\linewidth}{!}{    \begin{tabular}{lccrr}
        \toprule
        \multicolumn{1}{c}{\textbf{Methods}} & \textbf{Blind-spot} & \textbf{Train Cost} & \textbf{SIDD Val} & \textbf{SIDD Bench} \\
        \midrule
        Supervised &  no & $\times 1$  & 51.19/0.991 & 50.61/0.991 \\
         Noise2Noise &  no & $\times 1$  & 51.21/0.991 & 50.62/0.991 \\
        \midrule
        Noisy-as-Clean (NAC) &  no & $\times 1$  & 43.24/0.961 & - \\
        R2R         &  no  & $\times 1$ & 47.20/0.980 & 46.70/0.978  \\
        Noise2Void (N2V) & yes  & $\times 1$ & 48.55/0.984 & 48.01/0.983 \\
        NBR2NBR      & no  & $\times 2$ & 51.06/0.991 & 50.47/0.990 \\
        DCD-Net  & yes & $\times 3 $ & \underline{51.40}/\textbf{0.992} & - \\ 
        Blind2Sound (B2S) & yes  & $\times 4$ & \textbf{51.50}/\textbf{0.992} & \textbf{ 50.92}/\textbf{0.991} \\
        Blind2Unblind (B2U) & yes  & $\times 17$ & 51.36/\textbf{0.992} & \underline{50.79}/\textbf{0.991} \\
        L2R (Ours) &  no  & $\times 1.4$ & 51.35/\textbf{0.992}  & 50.60/\textbf{0.991}\\
        \bottomrule
    \end{tabular} } 
\end{table}

\subsection*{Real Cryo-EM Data Experiment.}

We evaluate the proposed L2R method on real Cryo-EM data from the
Topaz-EM open-source library~\citep{bepler2020topaz}, for which the noise
distribution is unknown. The dataset comprises five images with a
resolution of \(7676 \times 7420\) pixels. We employ the
local signal-dependent recorruptor from Eq.~\eqref{eqn:signaldepent}. A
single denoiser--recorruptor pair is trained jointly across all five
images, with the monotonic mapping and learned noise statistics shared
across images. These statistics are learned directly from the noisy
observations without prior knowledge of the noise parameters. For
comparison, we include results obtained with the PG-UNSURE approach
introduced by \citet{tachellaunsure}. Both methods are instantiated using a
four-scale bias-free U-Net. Training is performed using random
\(512 \times 512\) crops and the Adam optimizer, with a learning rate
of \(1 \times 10^{-4}\) and otherwise standard hyperparameters. Both
methods produce visually plausible denoising results, but L2R
produces smoother background textures than PG-UNSURE
while preserving visible particle structures.
Unlike PG-UNSURE, L2R does not require additional denoiser
evaluations.

\begin{figure}[!htbp]
    \centering
    \includegraphics[width=\linewidth]{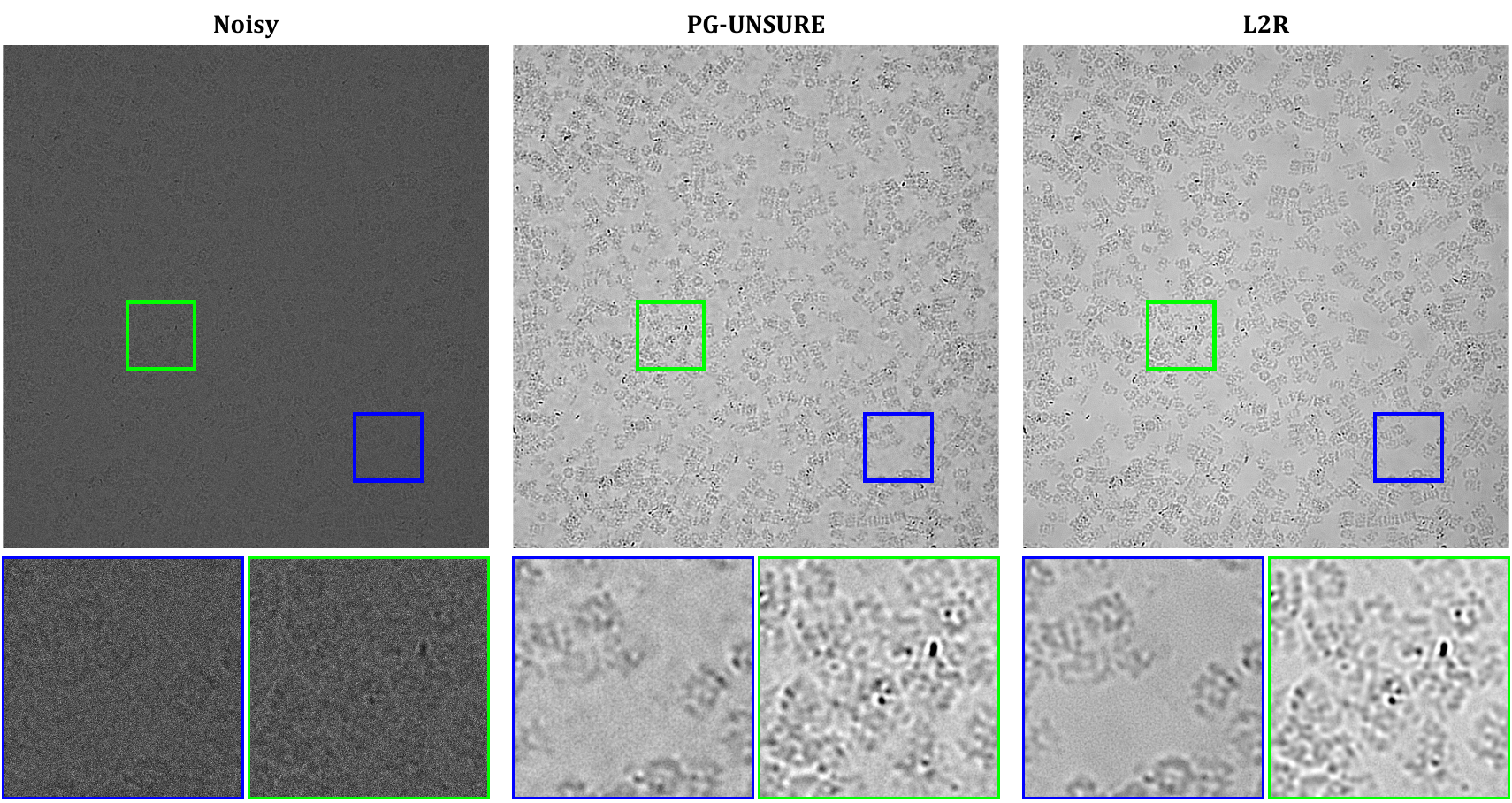}
    \caption{Blind denoising results of Cryo-EM data using the L2R method and PG-UNSURE.}
    \label{fig:cryoem}
\end{figure}

\section{Conclusions}
This work introduced \emph{Learning to Recorrupt} (L2R), a unified self-supervised framework combining a noise distribution agnostic objective with a learnable recorruptor. L2R learns the auxiliary GR2R statistics through a min--max objective without requiring explicit noise likelihoods, known statistics, or a moment-matched recorruption distribution a priori. Using the same objective and denoiser while adapting only the recorruptor structure, L2R achieves the highest PSNR among distribution-agnostic methods in 11 out of 12 non-Gaussian settings. On real SIDD data, it matches the Supervised and Noise2Noise references within \(0.02\) dB while remaining competitive with specialized blind-spot methods at substantially lower training cost. These results demonstrate the method's effectiveness across controlled and real-world noise. Beyond denoising, the learned recorruptor provides a proxy for the effective auxiliary noise statistics. Future work will explore more general recorruptor architectures for independent, correlated, and signal-dependent noise, as well as extensions of this approach to broader inverse problems with unknown measurement noise.

\section*{AI Use Statement}
In this work, we used generative AI tools to assist with method implementation and translation. We did not use generative AI tools to generate synthetic datasets, help develop theoretical models or conceptual frameworks, formulate mathematical claims, provide critical ingredients for proving mathematical claims, assist in writing proofs, propose or refine hypotheses, design or provide feedback on research methodology or experiments, clean or reformat datasets, support qualitative and thematic data analysis, or interpret results; these uses were not applicable to this work. Additionally, we used generative AI tools to create or modify scientific figures or images, create or edit software code, and summarize or analyze existing literature. We reviewed all AI-assisted work, and LLM-generated code was verified and tested for correctness by the authors. We take responsibility for the final content of this work, including text, claims, and artifacts produced with the aid of generative AI.

\section*{Acknowledgements}

This work is supported by the NVIDIA Academic Grant Program, which includes A100 80 GB GPU hours of computing and NVIDIA's Brev platform. J. Tachella acknowledges support from the ANR grant UNLIP 23-CE23-0013. Additionally, J. Tachella would like to thank Loic Denis for interesting discussions on the topic.

{
    \small
    \setlength{\parskip}{0pt}
    \setlength{\bibsep}{1pt}
    \bibliographystyle{iclr2027_conference}
    \bibliography{main}
}

\break 

\appendix
\section{Supplementary Material}

\subsection{Connection to Supervised Denoising} 
\label{sec:sm_eq16_eq17}

This section examines the connection between the proposed self-supervised constrained optimization and supervised denoising. We focus on the role of an auxiliary recorruption operator $h$, which generates the perturbation $h(\w')$ used to construct the recorrupted observation $\y_1=\y+\tau h(\w')$. Following Proposition~1 of \citet{monroy2025generalized}, we consider a candidate operator $h_c$ whose auxiliary noise $\boldsymbol{\omega}=h_c(\w')$ satisfies the mixed-moment conditions relevant to the denoiser class. For i.i.d.\ additive noise, these conditions are
\begin{equation}
    \mathbb{E}_{\varepsilon_i,\omega_i}
    \left[
    \left(\varepsilon_i-\frac{\omega_i}{\tau}\right)
    \left(\varepsilon_i+\tau\omega_i\right)^k
    \right] = 0,
    \qquad i=1,\ldots,n,\quad k=0,\ldots,K,
    \label{eq:sm_compatibility}
\end{equation}
where $\varepsilon_i$ is the $i$th component of the measurement noise and $\omega_i$ is the $i$th component of $h_c(\w')$. For spatially correlated noise, the corresponding conditions are expressed through the appropriate mixed moments. These conditions characterize compatibility at the level of the required moments; they do not require the auxiliary perturbation to reproduce the measurement noise sample by sample. Consider the noisy image $\y$ and recorrupted image $\y_1$
\begin{equation}
    \y = \x + g(\w), \qquad \y_1 = \y + \tau h(\w'),
\end{equation}
where \(\w,\w' \sim \mathcal{N}(\mathbf{0},\mathbf{I})\) are independent and \(\bvar = g(\w)\). Then
\begin{equation}
    \y_1 = \x + g(\w) + \tau h(\w').
\end{equation}
Using $\y=\x+g(\w)$, the self-supervised loss can be expanded as
\begin{equation}
\begin{aligned}
    \mathbb{E}_{\y_1,\y}\|f(\y_1)-\y\|_2^2
    &= \mathbb{E}_{\x,\w,\w'}\|f(\x+g(\w)+\tau h(\w'))-(\x+g(\w))\|_2^2 \\
    &= \mathbb{E}_{\x,\w,\w'}\|f(\x+g(\w)+\tau h(\w'))-\x\|_2^2 \\
    &\quad - 2\,\mathbb{E}_{\x,\w,\w'} f(\x+g(\w)+\tau h(\w'))^\top g(\w)
    + \text{const.}
\end{aligned}
\label{eq:sm_selfsup_decomp}
\end{equation}
The term denoted by \texttt{const.} is independent of $f$. On the other hand, the proposed method imposes that, for every admissible recorruption map $h\in\mathcal{H}$,
\begin{equation}
    \mathbb{E}_{\y,\w'} f(\y+\tau h(\w'))^\top h(\w') = 0.
    \label{eq:sm_constraint_all_h}
\end{equation}
Since $\y=\x+g(\w)$, this can be rewritten as
\begin{equation}
    \mathbb{E}_{\x,\w,\w'} f(\x+g(\w)+\tau h(\w'))^\top h(\w') = 0,
    \qquad \forall h\in\mathcal{H}.
    \label{eq:sm_constraint_rewritten}
\end{equation}
Thus, the feasible set consists of denoisers whose outputs are orthogonal in expectation to every admissible recorruption direction generated by $\mathcal{H}$. Now assume that $\mathcal{H}$ contains a compatible auxiliary operator $h_c$ satisfying Eq.~\eqref{eq:sm_compatibility} for the selected $\tau$ and the considered denoiser class. Since the constraint in Eq.~\eqref{eq:sm_constraint_rewritten} must hold for all $h\in\mathcal{H}$, it must in particular hold for the specific choice $h=h_c$, which yields
\begin{equation}
    {
    \mathbb{E}_{\x,\w,\w'}
    f(\x+g(\w)+\tau h_c(\w'))^\top h_c(\w') = 0.
    }
    \label{eq:sm_constraint_h_equals_g}
\end{equation}
For this compatible operator, the moment relations imply the following correlation identity~\citep{pang2021recorrupted}:
\begin{equation}
    {
    \frac{1}{\tau}\,
    \mathbb{E}_{\x,\w,\w'}
    f(\x+g(\w)+\tau h_c(\w'))^\top h_c(\w')
    =
    \mathbb{E}_{\x,\w,\w'}
    f(\x+g(\w)+\tau h_c(\w'))^\top g(\w).
    }
    \label{eq:sm_exchangeability}
\end{equation}
Hence, the constraint associated with $h_c$ also makes the correlation term with the true noise vanish. For any feasible $f$, the decomposition in Eq.~\eqref{eq:sm_selfsup_decomp} together with the previous argument then yields the following supervised-risk correspondence. Over the feasible set,
\begin{equation}
    \mathbb{E}_{\y_1,\y}\|f(\y_1)-\y\|_2^2
    =
    \mathbb{E}_{\y_1,\x}\|f(\y_1)-\x\|_2^2 + \text{const.}
\end{equation}
Since the additive constant does not depend on $f$, minimizing the self-supervised objective over the constraint set 
\begin{equation}
    \arg\min_f \; \mathbb{E}_{\y_1,\y}\|f(\y_1)-\y\|^2_2
    \quad \text{s.t.} \quad
    \mathbb{E}_{\y,\w'}f(\y+\tau h(\w'))^{\top} h(\w') = 0
    \quad \forall h \in \mathcal{H},
\end{equation}
can be written in the following supervised-risk form over the same constraint set:
\begin{equation}
    \arg\min_f \; \mathbb{E}_{\y_1,\x}\|f(\y_1)-\x\|^2_2
    \quad \text{s.t.} \quad
    \mathbb{E}_{\y,\w'}f(\y+\tau h(\w'))^{\top} h(\w') = 0
    \quad \forall h \in \mathcal{H}.
\end{equation}

\paragraph{Interpretation.}
The central idea is that the constraint family indexed by $h\in\mathcal{H}$ should be expressive enough to contain a compatible auxiliary operator satisfying the moment relations in Eq.~\eqref{eq:sm_compatibility}. In that case, the feasible set excludes denoisers whose outputs remain correlated with the true corruption, even though $g$ is never directly observed. The preceding derivation therefore provides a formal connection between the self-supervised objective and the supervised denoising risk, up to a term that does not depend on the denoiser parameters. During training, the maximization over the operator $h$ searches within $\mathcal{H}$ for auxiliary directions that enforce these constraints, while the denoiser minimizes the resulting objective. If $\mathcal{H}$ contains such a compatible operator and the saddle-point optimization is successful, the self-supervised objective can serve as a supervised-risk surrogate under the compatibility assumption; when the required moment relations cannot be represented exactly, this correspondence should be understood as approximate, which motivates choosing a sufficiently expressive class $\mathcal{H}$ for the learned recorruption operator.

\subsubsection{Relationship to GR2R}\label{sec:gr2r_relation}

The connection with GR2R becomes clearer by introducing, as in the main text, the paired observation
\begin{equation}
    \y_1 = \y + \tau h(\w'), \qquad
    \y_2 = \y - \frac{1}{\tau} h(\w'),
    \label{eq:sm_l2r_gr2r_pair}
\end{equation}
where in the present setting the recorruption map $h$ is \emph{learned} rather than fixed from a known noise model. Using this notation, the GR2R-type loss takes the form
\begin{equation}
    \mathcal{L}_{\mathrm{GR2R}}(\y;f) = \mathbb{E}_{\y_1,\y_2|\y} \|f(\y_1)-\y_2\|_2^2.
\end{equation}
Expanding $\y_2=\y-\frac{1}{\tau}h(\w')$ yields
\begin{equation}
\begin{aligned}
    \|f(\y_1)-\y_2\|_2^2
    &= \left\|f(\y_1)-\y+\frac{1}{\tau}h(\w')\right\|_2^2 \\
    &= \|f(\y_1)-\y\|_2^2
    + \frac{2}{\tau}\big(f(\y_1)-\y\big)^\top h(\w')
    + \frac{1}{\tau^2}\|h(\w')\|_2^2 \\
    &= \|f(\y_1)-\y\|_2^2
    + \frac{2}{\tau} f(\y_1)^\top h(\w')
    - \frac{2}{\tau}\y^\top h(\w')
    + \frac{1}{\tau^2} h(\w')^\top h(\w').
\end{aligned}
\label{eq:sm_gr2r_expansion_1}
\end{equation}
Rearranging the last two terms and using the definition of $\y_2$, one obtains
\begin{equation}
\begin{aligned}
    \|f(\y_1)-\y_2\|_2^2
    &= \|f(\y_1)-\y\|_2^2
    + \frac{2}{\tau} f(\y_1)^\top h(\w')
    - \frac{1}{\tau}(\y+\y_2)^\top h(\w').
\end{aligned}
\label{eq:sm_gr2r_expansion_2}
\end{equation}
Taking expectations, and recalling the proposed min--max objective
\begin{equation}
\min_f \max_h \;
\mathbb{E}_{\y,\w'}\Big[
    \|f(\y_1)-\y\|_2^2
    + \frac{2}{\tau} f(\y_1)^\top h(\w')
\Big],
\label{eq:sm_lagrange_recalled}
\end{equation}
the method can be interpreted as a GR2R-type criterion in which the paired recorruption process is learned adversarially:
\begin{equation}
\begin{aligned}
    \min_f \max_h \;
    \mathbb{E}_{\y,\w'}\Big[
        \|f(\y_1)-\y\|_2^2
        + \frac{2}{\tau} f(\y_1)^\top h(\w')
    \Big]
    \\ =
    \min_f \max_h \;
    \mathbb{E}_{\y_1,\y_2}\|f(\y_1)-\y_2\|_2^2
    + \mathbb{E}_{\y,\w'} \frac{1}{\tau}(\y+\y_2)^\top h(\w').
\end{aligned}
\label{eq:sm_l2r_as_gr2r}
\end{equation}
This expression highlights the main distinction with standard GR2R. In GR2R, the paired variables $(\y_1,\y_2)$ are generated from a \emph{known} oracle recorruption mechanism, so the loss $\mathbb{E}\|f(\y_1)-\y_2\|_2^2$ is optimized only with respect to $f$, and all remaining terms are fixed by construction. In the proposed setting, however, $h$ is itself an optimization variable. As a consequence, the additional term
\begin{equation}
    - \frac{1}{\tau}(\y+\y_2)^\top h(\w')
\end{equation}
cannot be ignored when optimizing with respect to $h$, because it induces an extra bias on the learned recorruption map. This term is irrelevant for the update of $f$, since it does not depend on the denoiser parameters, but it does affect the update of $h$ and may drive the recorruption mechanism toward degenerate solutions unrelated to the desired orthogonality condition. 

Hence, the proposed approach can be understood as a learned extension of GR2R in which the oracle recorruption process is no longer assumed known and is instead approximated through the min--max optimization; see Appendix~\ref{sec:optimsetup}. In this view, GR2R corresponds to the ideal case where the correct recorruption map is available from the outset, whereas the present method gradually guides $h$ toward an oracle GR2R-like mechanism while simultaneously training the denoiser. The optimization encourages $h$ to capture auxiliary statistics compatible with GR2R.

\subsubsection{Relationship to UNSURE}\label{sec:unsure_relation}

The proposed objective can also be interpreted as a learned generalization of UNSURE. Recall the L2R saddle objective
\begin{equation}
\min_f \max_h \;
\mathbb{E}_{\y,\w'}\Big[
    \|f(\y+\tau h(\w'))-\y\|_2^2
    + \frac{2}{\tau} f(\y+\tau h(\w'))^\top h(\w')
\Big].
\label{eq:sm_l2r_objective_unsure}
\end{equation}
Since, by construction, we ensure that $\mathbb{E}_{\w'}[h(\w')]=\mathbf{0}$, the correlation term can be equivalently rewritten as
\begin{equation}
\begin{aligned}
    \frac{2}{\tau}\mathbb{E}_{\y,\w'}\!\left[f(\y+\tau h(\w'))^\top h(\w')\right]
    &=
    \frac{2}{\tau}\mathbb{E}_{\y,\w'}\!\left[\big(f(\y+\tau h(\w'))-f(\y)\big)^\top h(\w')\right],
\end{aligned}
\label{eq:sm_l2r_unsure_centered}
\end{equation}
where the subtraction of $f(\y)$ is introduced only to center the perturbation term and expose a finite-difference form. This rewriting is useful because the quantity
\begin{equation}
    \frac{f(\y+\tau h(\w'))-f(\y)}{\tau}
\end{equation}
is precisely the first-order increment of $f$ along the perturbation direction $h(\w')$. Assuming that $f$ is differentiable and $\tau$ is small, a first-order expansion gives
\begin{equation}
    f(\y+\tau h(\w')) = f(\y) + \tau \frac{\partial f}{\partial \y}(\y)\, h(\w') + \mathcal{O}(\tau^2),
\end{equation}
and therefore
\begin{equation}
\begin{aligned}
    \frac{2}{\tau}\mathbb{E}_{\y,\w'}\!\left[f(\y+\tau h(\w'))^\top h(\w')\right]
    &=
    2\,\mathbb{E}_{\y}\!\left[
        \mathrm{tr}\!\left(
            \mathbf{C}_{h}\,\frac{\partial f}{\partial \y}(\y)
        \right)
    \right]
    + \mathcal{O}(\tau),
\end{aligned}
\label{eq:sm_l2r_unsure_general}
\end{equation}
where
\begin{equation}
    \mathbf{C}_{h} \triangleq \mathbb{E}_{\w'}\!\left[h(\w')h(\w')^\top\right].
\end{equation}
For the quadratic term, the same first-order expansion implies
\begin{equation}
\begin{aligned}
    \mathbb{E}_{\y,\w'}\|f(\y+\tau h(\w'))-\y\|_2^2
    =
    \mathbb{E}_{\y}\|f(\y)-\y\|_2^2
    + \mathcal{O}(\tau).
\end{aligned}
\end{equation}
Hence, to first order in $\tau$, the proposed loss becomes
\begin{equation}
\begin{aligned}
    \mathbb{E}_{\y}\|f(\y)-\y\|_2^2
    + 2\,\mathbb{E}_{\y}\!\left[
        \mathrm{tr}\!\left(
            \mathbf{C}_{h}\,\frac{\partial f}{\partial \y}(\y)
        \right)
    \right]
    + \mathcal{O}(\tau),
\end{aligned}
\label{eq:sm_l2r_unsure_master}
\end{equation}
which recovers several UNSURE-type objectives as special cases. From a recorruption-based perspective, L2R involves fewer approximations than UNSURE: it optimizes the recorrupted-input $\y_1$ objective directly, whereas the UNSURE connection emerges after a first-order reduction in which the correlation term is converted into a divergence penalty and the quadratic term is replaced by its small-$\tau$ surrogate. In addition, UNSURE requires a Monte Carlo finite-difference approximation of the divergence in practice, while L2R avoids divergence estimation altogether.

\noindent\textbf{Three cases of the L2R--UNSURE connection.}
We consider three noise settings to show how the parameterization of the L2R recorruptor $h$ shapes its connection to UNSURE. For isotropic and correlated Gaussian noise, linear recorruptors lead to the corresponding UNSURE divergence terms in the small-$\tau$ limit. For heavy-tailed noise, a nonlinear recorruptor retains higher-order contributions beyond that first-order limit.

\paragraph{1) Isotropic Gaussian noise.}
For the basic Gaussian denoising setting, choose a linear scaling recorruptor
\begin{equation}
    h(\w') = \sqrt{\alpha}\,\w',
    \qquad \w' \sim \mathcal{N}(\mathbf{0},\mathbf{I}),
\end{equation}
so that $\mathbf{C}_{h}=\alpha \mathbf{I}$. Then Eq.~\eqref{eq:sm_l2r_unsure_master} reduces to
\begin{equation}
    \mathbb{E}_{\y}\|f(\y)-\y\|_2^2
    + 2\alpha\,\mathbb{E}_{\y}\!\left[\mathrm{div}\,f(\y)\right]
    + \mathcal{O}(\tau),
\end{equation}
For a scalar linear-Gaussian recorruptor, this has the UNSURE form after identifying the UNSURE multiplier with the auxiliary covariance parameter, $\eta=\alpha$; the remaining discrepancy is $\mathcal{O}(\tau)$. Thus, as $\tau\to 0$, L2R recovers the UNSURE objective.

\paragraph{2) Correlated Gaussian noise.}
For correlated Gaussian noise, let the recorruptor be a linear operator applied to white Gaussian noise,
\begin{equation}
    h(\w') = \mathbf{H}_{\alpha}\w',
\end{equation}
or, equivalently in the spatially stationary case,
\begin{equation}
    h(\w') = k_{\alpha} * \w',
\end{equation}
with $*$ denoting convolution. In this case,
\begin{equation}
    \mathbf{C}_{h}
    =
    \mathbb{E}_{\w'}[\mathbf{H}_{\alpha}\w' \w'^{\top} \mathbf{H}_{\alpha}^\top]
    =
    \mathbf{H}_{\alpha}\mathbf{H}_{\alpha}^\top
    \triangleq \mathbf{\Sigma}_{\alpha},
\end{equation}
and Eq.~\eqref{eq:sm_l2r_unsure_master} becomes
\begin{equation}
    \mathbb{E}_{\y}\|f(\y)-\y\|_2^2
    + 2\,\mathbb{E}_{\y}\!\left[
        \mathrm{tr}\!\left(
            \mathbf{\Sigma}_{\alpha}\,\frac{\partial f}{\partial \y}(\y)
        \right)
    \right]
    + \mathcal{O}(\tau).
\end{equation}
This is the correlated-Gaussian version of UNSURE. In particular, when $\mathbf{H}_{\alpha}$ is chosen as a circulant convolution operator, the induced covariance $\mathbf{\Sigma}_{\alpha}$ coincides with the convolutional parameterization used by correlated-noise UNSURE (C-UNSURE)~\citep{tachellaunsure}. Hence, restricting $h$ to a linear convolution map recovers the same family of covariance-weighted divergence constraints.

\paragraph{3) Heavy-tailed noise.}
The previous two cases show that UNSURE is recovered when $h$ is restricted to linear transformations of Gaussian noise. However, for heavier-tailed corruptions such as log-gamma and Laplace, the proposed formulation goes beyond the basic Gaussian and correlated-Gaussian UNSURE objectives. Indeed, the heavy-tailed UNSURE formulation still imposes a structured constraint of the form
\begin{equation}
    \min_f \max_{\eta}\;
    \mathbb{E}_{\y}\|f(\y)-\y\|_2^2
    + 2\eta \sum_{i=1}^{n}
    \mathbb{E}_{\y}\!\left[
        a(y_i)\,\frac{\partial f_i}{\partial y_i}(\y)
    \right],
\end{equation}
where the weighting function $a(\cdot)$ must be specified analytically. In contrast, L2R replaces this hand-crafted scalar weighting by a learned nonlinear recorruption map $h\in\mathcal{H}$ and enforces the wider family of constraints
\begin{equation}
    \mathbb{E}_{\y,\w'}\!\left[
        f(\y+\tau h(\w'))^\top h(\w')
    \right] = 0,
    \qquad \forall h \in \mathcal{H}.
\end{equation}
The additional flexibility appears beyond the first-order UNSURE limit. In scalar notation, with $h=h(\w')$, retaining higher-order terms gives
\begin{equation}
\begin{aligned}
    \frac{2}{\tau}\mathbb{E}_{\y,\w'}[f(\y+\tau h)h]
    &=2\mathbb{E}_{\y}[f'(\y)]\mathbb{E}_{\w'}[h^2]
    +\tau\mathbb{E}_{\y}[f''(\y)]\mathbb{E}_{\w'}[h^3] \\
    &\quad+\frac{\tau^2}{3}\mathbb{E}_{\y}[f'''(\y)]\mathbb{E}_{\w'}[h^4]
    +\mathcal{O}(\tau^3),
\end{aligned}
\label{eq:sm_higher_order_unsure}
\end{equation}
showing that finite-$\tau$ nonlinear recorruption incorporates variance,
skewness, and higher-order moments. Therefore, UNSURE is recovered under a
linear first-order parameterization, whereas the learned monotonic recorruptor
defines a richer family that is particularly relevant for heavy-tailed noise.
L2R also avoids explicit divergence estimation and the additional denoiser
evaluations used by Monte Carlo finite differences. This interpretation agrees
with the empirical moment matching and the improvements over UNSURE reported
for log-gamma and Laplace noise.

\subsection{Implementation Details}

\subsubsection{Noise models and recorruptor parameterization}
\label{sec:noisemodels}

We evaluate performance under a diverse set of non-Gaussian noise regimes that capture statistical properties observed in real data, including heavy-tailed log-gamma, Laplace, spatially correlated Gaussian, and Poisson--Gaussian noise, in order to stress-test robustness beyond the standard i.i.d. Gaussian assumption.

\begin{figure}[h]
    \centering
    \includegraphics[width=\linewidth]{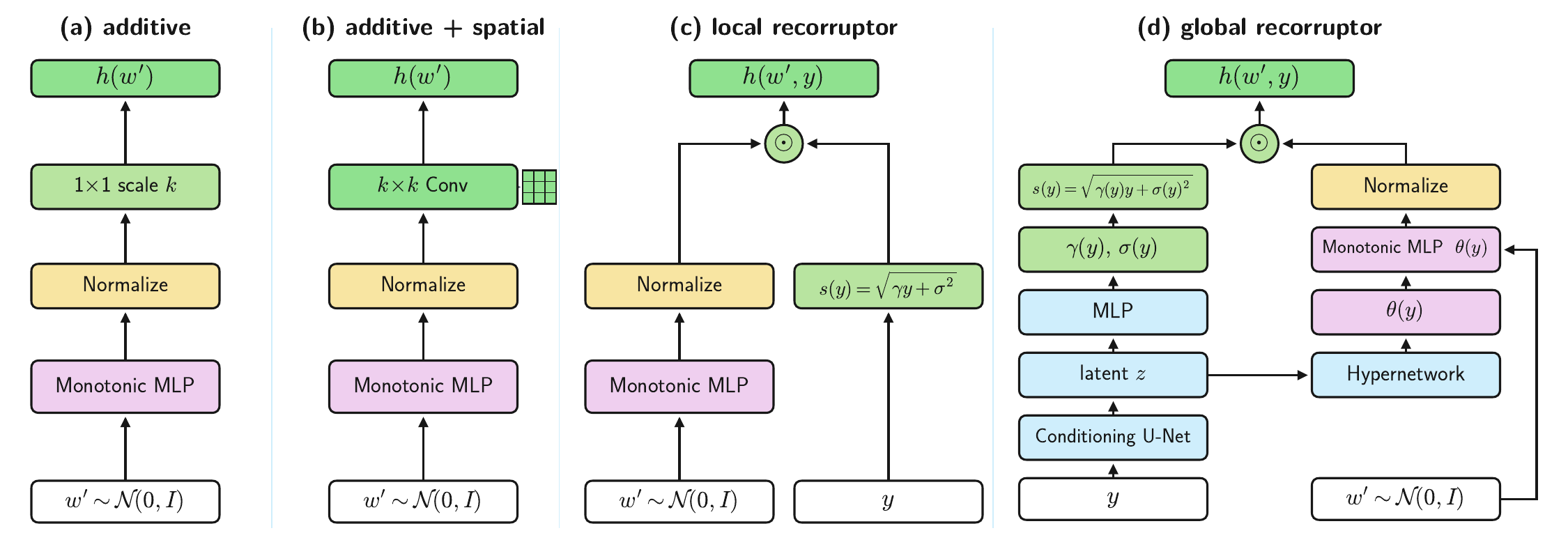}
    \caption{\textbf{Overview of the four recorruptor parameterizations.} (a) Additive i.i.d. noise; (b) additive noise with spatial correlations; (c) local recorruptor, with parameters shared across images and pixelwise scaling from $\y$; and (d) global recorruptor, which uses image-level information from $\y$ to estimate image-specific noise statistics and monotonic-map parameters.}
    \label{fig:recorruptor}
\end{figure}

\paragraph{Log-gamma noise.} We model zero-mean additive log-gamma noise with fixed variance to capture different long-tailed distributions. Specifically, we parametrize the noise variable as $\boldsymbol{\varepsilon} = \tilde{\sigma}(\ln(\boldsymbol{z}) - c_\ell)$, where $\boldsymbol{z}$ follows a Gamma distribution $z_i\sim \mathcal{G}(\ell, \ell)$, with $\ell$ controlling the shape of the distribution. Here, $c_\ell$ removes the constant bias and $\tilde{\sigma}$ rescales to the target standard deviation. They can be computed as $c_\ell = \psi(\ell) - \ln(\ell)$ and $\tilde{\sigma} = \sigma /\sqrt{ \psi_1(\ell) } $, where $\psi$ and $\psi_1$ denote the digamma and trigamma functions, respectively. The standard deviation is fixed at \(\sigma = 0.1\), and the recorruptor uses a $1\times1$ kernel for element-wise recorruption.

\paragraph{Laplace noise.} In the case of additive Laplace noise, the noise variable is modeled as $\varepsilon_i \sim \text{Laplace}(0,b)$, where the location parameter is set to zero to ensure a zero mean, and the scale parameter is $b$. The corresponding standard deviation of the noise is given by $\sigma = b\sqrt{2}$. In this setting, the same element-wise recorruptor $h$ is employed.

\paragraph{Correlated noise.} In the correlated case, the noise variable is reparameterized as $\bvar = \boldsymbol{z} * \boldsymbol{k}_{\mathrm{gt}}$, where $\boldsymbol{z} \sim \mathcal{N}(\mathbf{0}, \sigma^2 \mathbf{I})$ and $\boldsymbol{k}_{\mathrm{gt}}$ denotes the ground-truth Gaussian blur kernel. In these experiments, $\boldsymbol{k}_{\mathrm{gt}}$ is instantiated with a spatial size $3 \times 3$ and a fixed standard deviation. The ground-truth kernel is presented in Figure~\ref{fig:mapping}. The recorruptor also uses a $3 \times 3$ kernel.

\paragraph{Poisson--Gaussian noise.}
Under the Poisson--Gaussian model,
\(\y=\gamma\boldsymbol{z}_p+\boldsymbol{z}\), with
\(\boldsymbol{z}_p\sim\mathcal{P}(\x/\gamma)\) and
\(\boldsymbol{z}\sim\mathcal{N}(\mathbf{0},\sigma^2\mathbf{I})\).
Hence,
\(\mathbb{E}[\y\mid\x]=\x\) and
\(\operatorname{Var}(\y\mid\x)=\gamma\x+\sigma^2\).
Equivalently,
\(\y=\x+\boldsymbol{\varepsilon}_{\x}\) with
\(\mathbb{E}[\boldsymbol{\varepsilon}_{\x}\mid\x]=\mathbf{0}\), so the GR2R moment relations extend by conditioning on \(\x\).
Following the conditional recorruptor defined above, we set
\(s(\y)=\sqrt{\gamma(\y)\y+\sigma(\y)^2}\), so that
\[
    h(\w', \y)
    =
    s(\y)\odot \tilde h_{\y}(\w').
\]
Here, \(s(\y)\) sets the signal-dependent recorruption scale, and \(\tilde h_{\y}\) follows the architecture in Eq.~\ref{eq:recorruptor_additive}.

We consider two configurations.

\emph{Local recorruptor (Fig.~\ref{fig:recorruptor}(c)).}
For the synthetic Poisson--Gaussian and Cryo-EM experiments, \(\gamma\), \(\sigma\), and the weights of \(\tilde h_{\y}\) are learned and shared across images. The scale \(s(\y)=\sqrt{\gamma\y+\sigma^2}\) varies pixelwise with \(\y\), while the learned statistics and transport remain shared.

\emph{Global recorruptor (Fig.~\ref{fig:recorruptor}(d)).}
For the real-world SIDD experiments, a conditioning U-Net extracts an image-level latent representation from \(\y\). One part is passed through an MLP to predict image-specific parameters \(\gamma(\y)\) and \(\sigma(\y)\); the other is passed to a hypernetwork that predicts the image-specific weights of the monotonic network defining \(\tilde h_{\y}\). An element-wise absolute-value activation constrains these weights to be nonnegative and preserves the monotonicity of \(\tilde h_{\y}\). Thus, information from the whole image determines its noise statistics and recorruptor, while \(s(\y)\) provides pixelwise scaling.

Finally, the choice of recorruptor class reflects a trade-off between generality and specialization. A more expressive, general-purpose recorruptor could represent multiple noise distribution families within a single model, but it also enlarges the search space and may make learning the appropriate recorruption for each distribution more difficult. In contrast, a specialized recorruptor that incorporates known structural properties of the corruption process can reduce the learning burden and provide a more accurate parameterization. For example, prior knowledge about whether the noise is additive, signal-dependent, or spatially correlated can guide the design of the recorruptor class. Thus, the choice of recorruptor architecture acts as both a user-specified and structural prior: greater knowledge of the real noise process enables more targeted modeling, whereas limited knowledge motivates more general but potentially harder-to-optimize classes. Developing recorruptors that balance broad distributional coverage with reliable optimization is an important direction for future work.

\subsection{Additional Simulations}

\subsubsection{Poisson--Gaussian Denoising}

\begin{wraptable}{r}{0.47\columnwidth}
\vspace{-1\baselineskip}
\centering

\caption{Denoising performance in removing Poisson--Gaussian noise. \label{tab:poisson}} 
\resizebox{\linewidth}{!}{\begin{tabular}{lc|ccc} 
\toprule
 \multirow{2}{*}{\bb{Method}}
 &  $p_{\bvar}$
 & \multicolumn{3}{c}{\bb{Poisson--Gaussian Noise}} 
 \\ 
 & known? &  \quad$\gamma/\sigma$\quad    & \bb{BSDS500} & \bb{DIV2K} 
 \\ \midrule \midrule 
  \rowcolor{gray!15} 
   Supervised  
   &  no  &    &   28.44/0.833 & 29.45/0.845 
\\ 
 \rowcolor{gray!15} 
 PG-SURE    
 & yes  &    &  27.21/\underline{0.776} & 27.94/\underline{0.783}
  \\   PG-UNSURE   
 &  no  &  0.05/   & \underline{27.44}/0.774 &\underline{28.14}/0.771
 \\   NBR2NBR  
 &  no &  0.05\;  &  27.12/0.751 & 27.64/0.744  
 \\   Blind2Sound 
 &  no &  &  27.30/0.793 & 27.89/0.789
 \\   L2R (Ours)
 &  no &    &  \textbf{27.80/0.791} & \textbf{28.70/0.797}
 \\ \midrule   \rowcolor{gray!15}
    Supervised  
   &  no  &    &  27.16/0.792 & 28.27/0.815
 \\  \rowcolor{gray!15}
 GR2R
 &  yes &    &  27.03/0.786 & 28.15/0.812
 \\    \rowcolor{gray!15} 
 PG-SURE 
 &  yes &    &  25.91/0.729 & 26.69/0.744
 \\   PG-UNSURE
 &  no &  0.1/   &  25.80/0.710 & 26.57/0.724
 \\   NBR2NBR     
 &  no &  0.0\;  &  \underline{26.34/0.750} & \underline{27.20/0.768}  
 \\  Blind2Sound
 &  no &  &  26.12/0.747  & 26.78/0.754
 \\
   L2R (Ours) 
 &  no &    &  \textbf{26.63/0.756} & \textbf{27.68/0.779} 
 \\
                \bottomrule
\end{tabular} }
 \vspace{-1\baselineskip}
\end{wraptable}

We evaluate the proposed L2R under two Poisson--Gaussian denoising setups: a Poisson-only case ($\gamma/\sigma=0.1/0.0$) and a mixed Poisson--Gaussian case ($\gamma/\sigma=0.05/0.05$). Here, $\gamma$ controls the Poisson variance (gain) and $\sigma$ denotes the Gaussian standard deviation; in both setups the dominant corruption is still Poisson. Comparisons are conducted against representative self-supervised baselines, including PG-SURE and PG-UNSURE (the Poisson--Gaussian variants of SURE and UNSURE, respectively), as well as GR2R and NBR2NBR.

\begin{figure*}[h]
    \centering
    \includegraphics[width=\linewidth]{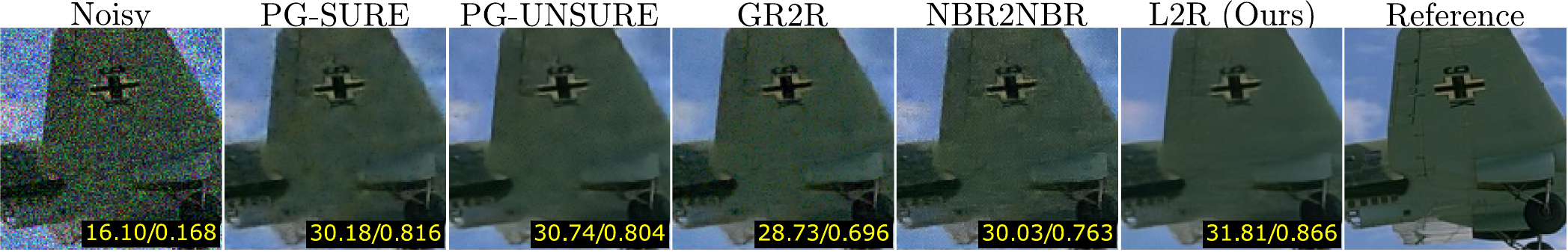}
    \includegraphics[width=\linewidth]{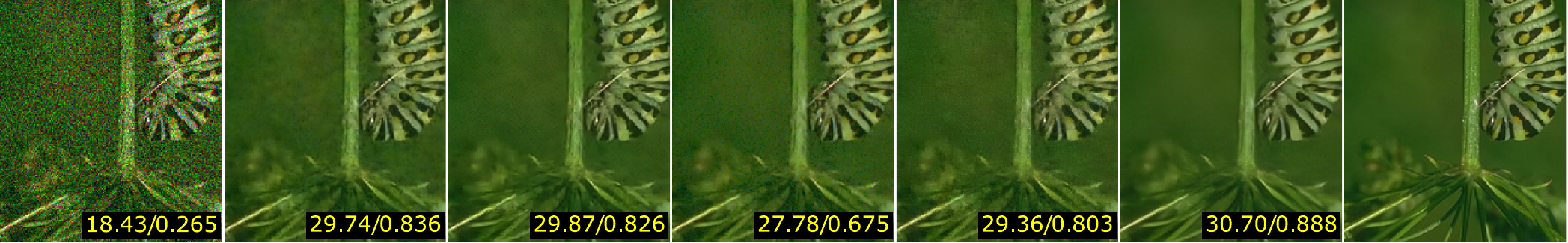} \vspace{-1.8em}
    \caption{\textbf{Poisson--Gaussian Denoising.} Visual comparison on representative BSDS500 crops corrupted by Poisson--Gaussian Noise. Columns show the noisy input, PG-SURE, PG-UNSURE, GR2R, NBR2NBR, and the proposed L2R, followed by the ground truth. }
    \label{fig:visualpoisson} \vspace{-1em}
\end{figure*}

Figure~\ref{fig:visualpoisson} shows that L2R yields cleaner reconstructions with improved edge fidelity and fewer residual artifacts across representative BSDS500 crops. Compared to the strongest and most methodologically related baseline, PG-UNSURE, L2R produces sharper structures and reduced noise remnants.

Table~\ref{tab:poisson} confirms these observations: L2R achieves the best PSNR among methods that do not assume knowledge of the noise distribution. At $\gamma/\sigma=0.05/0.05$, L2R improves over PG-UNSURE from 27.44 to 27.80~dB on BSDS500 and from 28.14 to 28.70~dB on DIV2K. Notably, PG-UNSURE relies on a Monte Carlo divergence approximation to build its objective, whereas L2R avoids divergence estimation, removing this additional approximation while still improving PSNR. In contrast, GR2R performs the best for Poisson-only denoising ($\gamma=0.1$, $\sigma=0.0$), given that it has exact prior knowledge of the noise distribution.

\pagebreak
\subsubsection{Monotonic learned recorruptor \(h\)}  

\begin{wraptable}{r}{0.45\columnwidth}
\vspace{-3\baselineskip}
\centering
\caption{Ablation of the design of model $h$.}
\label{tab:hifluence}  
\resizebox{\linewidth}{!}{\begin{tabular}{lll} \toprule
\textbf{network setup} $h(\cdot)$                      & \textbf{PSNR} & \textbf{SSIM} \\
\midrule
\rowcolor{gray!15} Supervised &  31.44  &  0.898     \\
\rowcolor{gray!15} Oracle (GR2R)  & 31.14 &  0.889     \\
No learning                   &  20.00  &  0.415    \\
MLP                           &  30.45  &  0.860     \\
MLP + Residual                &  30.36  &  0.858 \\
MLP + Id-pretrain          &  30.42  &  0.856 \\
Monotonic MLP                 &  30.45  &  0.864    \\
Monotonic MLP + Id-pretrain \quad\quad &  \textbf{30.77}  & \textbf{0.871}      \\   \bottomrule 
\end{tabular}
}

 \vspace{-0.8\baselineskip}
\end{wraptable}

Table~\ref{tab:hifluence} reports an ablation on the design of $h$ for self-supervised denoising of log-gamma noise with $\ell=1.0$. \emph{No learning} corresponds to evaluating the metrics directly on the noisy observation. Unconstrained MLP variants yield a large improvement and approach the GR2R oracle bound, but remain sensitive to optimization and do not explicitly enforce the intended order-preserving behavior. Adding a residual connection does not provide consistent gains. Enforcing monotonicity improves robustness, and combining it with identity initialization (\emph{Id-pretrain}) yields the best overall self-supervised performance. In particular, \emph{Monotonic MLP + Id-pretrain} achieves the top results (30.77\,dB / 0.871), narrowing the gap to supervised training (31.44\,dB / 0.898) while being competitive with GR2R in PSNR.

\subsubsection{Influence of training optimization setup \label{sec:optimsetup}}

Table~\ref{tab:jointloss} analyzes the effect of different optimization strategies for the denoiser $f$ and the learned recorruption map $h$ in the correlated Gaussian denoising setting with $\sigma=0.2$. The experiments preserve the same network, dataset, and training configuration reported in the main text. In the \emph{joint} setting, both variables are optimized from a single loss evaluation by applying gradient descent on $f$ and gradient ascent on $h$ using the same forward pass. In contrast, the \emph{non-joint} setting alternates the updates: the loss is first evaluated to update $f$, and then re-evaluated to update $h$. The results show that the two strategies perform similarly; the stop-gradient joint configuration gives the best result. This is relevant in practice, as the joint formulation is simpler and avoids repeated loss evaluations while preserving essentially the same denoising quality.

The table also compares GR2R-style and UNSURE-style formulations for the min--max objective. In particular, the GR2R-inspired versions $\Vert f(\y_1)-\y_2\Vert_2^2$ and their bias-corrected variant produce results that are close to those of the proposed UNSURE-style formulation, suggesting that both views are consistent in practice and that the empirical gap between them is minimal in this setting. However, the scaling of the correlation term is important: replacing the factor $\frac{2}{\tau}$ with an unscaled inner product $h(\w')^\top f(\y_1)$ consistently degrades performance, both in the joint and non-joint settings.

Finally, in the practical implementation we use a stop-gradient operation on $\y_1$ when updating the recorruptor, i.e., $f(\mathrm{sg}[\y_1])$. This yields the best overall results and prevents gradients from propagating through the denoiser with respect to the dependence of $\y_1$ on $h$. Consequently, the update of $h$ is driven directly by the correlation term, providing an efficient surrogate ascent step while avoiding an additional backward pass through $f$ and reducing computational overhead.

\renewcommand{\arraystretch}{1.5}
\begin{table}[h] 
\caption{\textbf{Ablation $f$-loss and $h$-loss.} Correlated denoising performance for different types of $f$-loss and $h$-loss with $\sigma=0.2$. $\text{sg}[\cdot]$ denotes the stop-gradient function. \label{tab:jointloss}}
\centering
\resizebox{0.7\linewidth}{!}{\begin{tabular}{l|l|c|cc} \toprule
Joint? &  $f$-loss (min) &   $h$-loss (max) &  PSNR & SSIM \\ \midrule
no &  $ \Vert f(\y_1) - \y_2 \Vert_2^2$     & $h(\w')^\top f(\y_1)$    &    29.72     &  0.854   \\
  no &     \multicolumn{2}{l|}{ $\Vert f(\y_1) - \y_2 \Vert_2^2 +  \frac{1}{\tau}\cdot(\y_2 + \y)^\top h(\w') $ }    &   29.74  &  0.854    \\
  no &     \multicolumn{2}{l|}{ $\Vert f(\y_1) - \y \Vert_2^2 \hspace{0.45em} + 2 \cdot h(\w')^\top f(\y_1)$ }    &   29.38  &   0.841      \\
 no &     \multicolumn{2}{l|}{ $\Vert f(\y_1) - \y \Vert_2^2 \hspace{0.45em} + \frac{2}{\tau}\cdot h(\w')^\top f(\y_1)$ }    &   29.83  &  0.859    \\  
 \midrule
     
 yes &     \multicolumn{2}{l|}{ $\Vert f(\y_1) - \y \Vert_2^2 \hspace{0.45em} + \frac{2}{\tau}\cdot h(\w')^\top f(\y_1)$ }   &   29.78  &   0.860      \\
 yes &     \multicolumn{2}{l|}{ $\Vert f(\text{sg}[\y_1]) - \y \Vert_2^2  + \frac{2}{\tau}\cdot h(\w')^\top f(\text{sg}[\y_1])$ }   &   29.85  &   0.863    
   \\  
 yes &     \multicolumn{2}{l|}{ $\Vert f(\text{sg}[\y_1]) - \y \Vert_2^2  + h(\w')^\top f(\text{sg}[\y_1])$ }   &   29.25  &   0.847   \\ 
  yes &     \multicolumn{2}{l|}{ $\Vert f(\y_1) - \y \Vert_2^2  + h(\w')^\top f(\y_1)$ }   &   28.85  &   0.825   \\ \bottomrule     
\end{tabular} 
} \vspace{-1em}
\end{table}

\subsubsection{Expressivity--Robustness Trade-off}

\begin{wraptable}{r}{0.45\columnwidth}
\footnotesize
\centering
\caption{Performance for different model sizes of $h$ on log-gamma noise. \label{tab:h_depth_width}}
\resizebox{\linewidth}{!}{\begin{tabular}{c|c|rr}
\toprule
\textbf{Depth} \quad &\;\#\textbf{Neurons} \quad &\quad \textbf{PSNR} &\quad  \textbf{SSIM} \\ \midrule
2     & 16        &    $30.57$ & $0.871$  \\
2     & 8         &    $30.60$  &  $0.874$     \\
3    & 8         &    $30.56$ &  $0.870$     \\
4    & 8         &    $30.53$ &  $0.871$     \\
2     & 4        &   $30.50$   &  $0.866$   \\
3     & 4        &    $30.60$   &  $0.872$   \\
4     & 4        &    $\underline{30.64}$   &  $\underline{0.875}$   \\
5     & 4        &   $\boldsymbol{30.79}$  &  $\boldsymbol{0.879}$  \\
6     & 4        &   $30.57$  &  $0.870$ \\
\bottomrule
\end{tabular}
}
\end{wraptable}

This section analyzes the expressivity--robustness trade-off of the learned recorruptor \(h\) through a denoising evaluation on log-gamma noise with \(\ell = 1.0\). In particular, the goal is to assess how the architectural capacity of \(h\), controlled by its depth and number of neurons per layer, affects the final reconstruction performance. Since \(h\) is trained adversarially to generate informative recorruptions, this experiment helps identify the model size that best balances representational flexibility and optimization robustness.

Table~\ref{tab:h_depth_width} shows how the design of \(h\) affects performance. The most relevant trend appears for the narrow configuration with \(4\) neurons per layer: increasing the depth from \(2\) to \(5\) improves the PSNR from \(30.50\) dB to \(30.79\) dB, with depth \(4\) already reaching \(30.64\) dB. This indicates that additional depth is beneficial because it increases the nonlinear expressivity of the recorruptor while keeping the number of parameters per layer controlled. In this case, depth allows \(h\) to better approximate the required recorruption mechanism without making each layer excessively flexible.

At the same time, the results show that increasing the width alone does not necessarily improve performance. For example, shallow but wider models such as depth \(2\) with \(8\) and \(16\) neurons achieve \(30.60\) dB and \(30.57\) dB, respectively, which remain below the deeper narrow model with depth \(5\) and width \(4\). Moreover, increasing depth too aggressively is also detrimental, since depth \(6\) with \(4\) neurons drops back to \(30.57\) dB.

\begin{table}[h] 
\centering
\caption{\textbf{Influence of the architecture design \(h\).} The table reports the ablation results for different kernel sizes and $s(\y)$-scaling configurations, evaluated under three noise models: element-wise log-gamma noise, spatially correlated noise, and signal-dependent Poisson--Gaussian noise. \label{tab:hdesign}}
\resizebox{0.55\linewidth}{!}{\begin{tabular}{l|cc|rr}
\toprule
\textbf{Noise Model} & \quad $k$\textbf{-size}\quad & \quad $s(\y)$  \quad & \textbf{PSNR} & \textbf{SSIM} \\
\midrule
Log-gamma \; 
& $1\times1$ & no & $\boldsymbol{30.77}$ & $\boldsymbol{0.871}$ \\
$\ell=1.0$
& $3\times3$ & no & $\underline{29.93}$ & $\underline{0.837}$ \\
& $1\times1$ & yes & $26.98$ & $0.721$ \\
& $3\times3$ & yes & $24.02$ & $0.657$ \\
\hline
Correlated \;
& $1\times1$ & no & $24.21$ & $0.660$ \\
$\sigma=0.2$ \;
& $3\times3$ & no & $\boldsymbol{29.85}$ & $\boldsymbol{0.863}$ \\
& $1\times1$ & yes & $24.02$ & $0.657$ \\
& $3\times3$ & yes & $\underline{27.05}$ & $\underline{0.754}$ \\
\hline
Poisson--Gaussian
& $1\times1$ & no & $24.99$ & $0.628$ \\
$\gamma=0.05,\sigma=0.05$ \;
& $3\times3$ & no & $23.77$ & $0.581 $ \\
& $1\times1$ & yes & $\boldsymbol{27.80}$ & $\boldsymbol{0.791}$ \\
& $3\times3$ & yes & $\underline{26.03}$ & $\underline{0.713}$ \\
\bottomrule
\end{tabular} 
}  
\end{table}

\subsubsection{Influence of the architecture design \(h\)}

Table~\ref{tab:hdesign} shows how the design of the recorruption mapping \(h\) affects performance across noise models, indicating that the best architecture depends on the corruption structure. For element-wise log-gamma noise, the pointwise \(1\times1\) design without $s(\y)$ scaling achieves the best performance, reaching \(30.77\) dB PSNR and \(0.871\) SSIM, outperforming both the \(3\times3\) variant and the scaled \(1\times1\) version. A similar tendency is observed for Poisson--Gaussian noise, where the \(1\times1\) architecture combined with $s(\y)$-scaling attains the best results (\(27.80\) dB, \(0.791\) SSIM), while removing the scaling leads to a noticeable degradation to \(24.99\) dB and \(0.628\) SSIM. 

In contrast, the correlated noise setting benefits from incorporating the spatial convolution operator in \(h\). In this case, the \(3\times3\) configuration without $s(\y)$ scaling reaches \(29.85\) dB and \(0.863\) SSIM, whereas the scaled \(3\times3\) version degrades to \(27.05\) dB and \(0.754\) SSIM, and the pointwise \(1\times1\) alternative further drops to \(24.21\) dB and \(0.660\) SSIM. This is expected, as employing a non-correlated recorruption scheme can induce substantial errors when the underlying noise exhibits spatial correlations.  Overall, the ablation indicates that the design of \(h\) should remain aligned with the main characteristics of the corruption process: element-wise mappings appear adequate for independent noise, spatial kernels become beneficial for correlated perturbations, and $s(\y)$ scaling is helpful in signal-dependent Poisson--Gaussian regimes.

\subsubsection{Training Stability}

We further evaluate the sensitivity of L2R to its main optimization
parameters by varying one parameter at a time while holding
the other settings fixed. The random seed is fixed within each
experiment.

Table~\ref{tab:tau} reports an additional sensitivity analysis of the
recorruption parameter $\tau$ under heavy-tailed log-gamma noise.
This experiment uses a separate random seed and 1,000 training
epochs. L2R exhibits a
broad effective operating region for moderate perturbation strengths,
with the strongest performance observed around $\tau\in[0.3,0.5]$.
In particular, $\tau=0.5$ attains the highest PSNR of $30.34$ dB
while essentially matching the best SSIM. Performance remains close
throughout $\tau\in[0.3,0.7]$, whereas more extreme values lead to
larger degradation. This indicates that $\tau$ controls a meaningful
recorruption trade-off. 
\begin{table}[htbp] \centering \footnotesize  
\caption{Sensitivity to $\tau$ \label{tab:tau} on log-gamma noise $\ell = 1.0$ } 
\begin{tabular}{lccccccc}
\toprule
$\tau$        & 0.1 & 0.3 & 0.4 & 0.5 & 0.6 & 0.7 & 0.9 \\ \midrule
\textbf{PSNR} & 28.83   &  \underline{30.21} &  29.95 &  \textbf{30.34}  & 29.78 &   29.59  &  29.26   \\
\textbf{SSIM} & 0.821   &  \textbf{0.869} & 0.865 &  \underline{0.868}   &  0.860 &   0.854  &  0.843 \\  \bottomrule
\end{tabular} 
\end{table}

We additionally report a separate SIDD experiment for sensitivity
analysis, using independent training runs; these values are not the
principal benchmark result in Table~\ref{tab:sensibilidad_5valores}.
Consistent with the scenario-dependent role of $\tau$, the favorable
operating region on SIDD is centered around $\tau=1$. Performance
remains stable in its neighborhood, varying by only $0.28$ dB for
$\tau\in[0.8,1.25]$, while more extreme perturbation strengths lead
to moderate degradation.

L2R is similarly robust to the relative learning rate between the
denoiser and recorruptor. Although $R_{lr}=0.5$ achieves the highest
PSNR of $51.32$ dB, values in the broad range
$R_{lr}\in[0.3,0.9]$ remain within $0.23$ dB of one another, with
$R_{lr}=0.9$ reaching $51.30$ dB. Finally, the method exhibits very
low sensitivity to stochastic initialization: five independent random
seeds yield $51.25\pm0.06$ dB, with all runs contained within a
$0.17$ dB interval. Overall, these results indicate that the min--max
optimization is reproducible across initializations and does not depend
on narrowly tuned optimization parameters.

\begin{table}[htbp] \centering \footnotesize
\caption{Training stability on SIDD under variations of the main optimization parameters.} \label{tab:sensibilidad_5valores} \renewcommand{\arraystretch}{1.0} \begin{tabular}{l c c} \toprule \textbf{Parameter} & \textbf{Values / PSNR (dB)} & \textbf{avg PSNR (mean $\pm$ std.)} \\ \midrule \textbf{Recorruption ($\tau$)} & $\begin{array}{c|c|c|c|c} \small 0.6 & \small 0.8 & \mathbf{1.0} & \small 1.25 & \small 1.67 \\ \hline \small 50.68 & \small 51.04 & \textbf{51.32} & \small 51.20 & \small 50.98 \end{array}$ & $51.04 \pm 0.24 $ \\ \addlinespace[0.6em] \textbf{Learning Ratio ($R_{lr}$)} & $\begin{array}{c|c|c|c|c} \small 0.1 & \small 0.3 & \mathbf{0.5} & \small 0.7 & \small 0.9 \\ \hline \small 50.76 & \small 51.09 & \textbf{51.32} & 51.14 & 51.30 \end{array}$ & $51.12 \pm 0.22 $ \\ \addlinespace[0.6em] \textbf{RNG Seed} & $\begin{array}{c|c|c|c|c} \small 0 & \small 1 & 2 & \small 3 & \small 4 \\ \hline 51.29 & 51.15 & \textbf{51.32} & 51.27 & 51.25 \end{array}$ & $51.25 \pm 0.06$ \\ \bottomrule \end{tabular} \end{table}

\subsection{Learned recorruption maps and spatial kernel}
\label{sec:implicit_moments_kernel}

\begin{figure}[!htbp]
    \centering
    \includegraphics[height=0.28\linewidth]{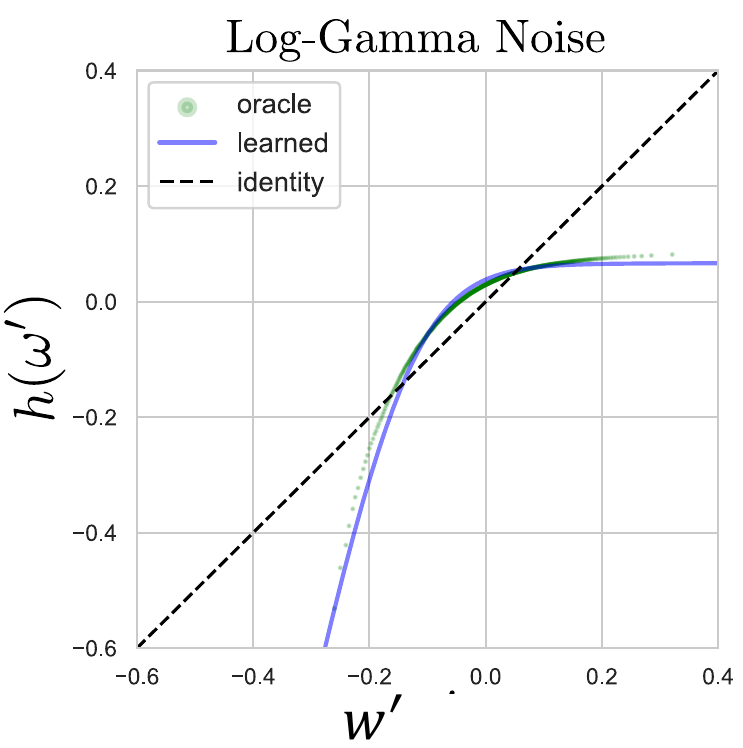}
     \includegraphics[height=0.28\linewidth]{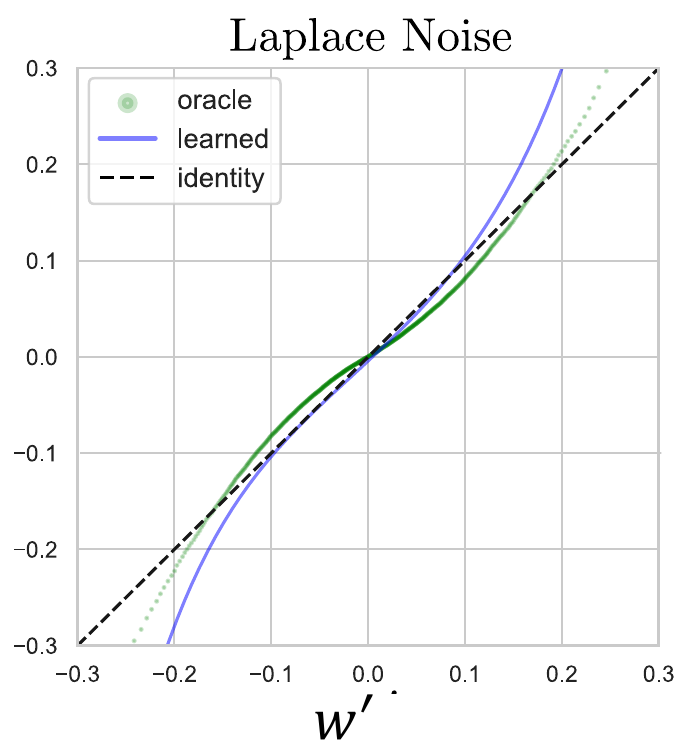}
    \includegraphics[height=0.28\linewidth]{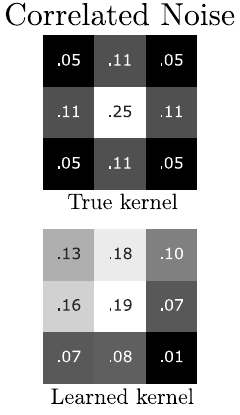}
    \vspace{-1em}
    \caption{\textbf{Learned maps and spatial kernel.} From \textit{left} to \textit{right}, the panels show the learned recorruption maps for log-gamma and Laplace noise and the learned kernel for correlated noise.}
    \label{fig:mapping}
\end{figure}

Figure~\ref{fig:mapping} compares the learned recorruption maps and spatial kernel with their oracle counterparts. For the log-gamma and Laplace settings (left and center), the recorruptor $h$ is compared with (i) an \emph{oracle} recorruption obtained by manually matching low-order moments of the target distribution, and (ii) the identity mapping, representing Gaussian noise. For correlated noise (right), the true $3\times 3$ Gaussian blur kernel inducing spatial correlation is compared with the kernel recovered by the proposed min--max training.

For log-gamma noise, the learned mapping closely follows the oracle curve over most of the support, exhibiting the expected nonlinearity required to reproduce the heavy-tailed behavior. For Laplace noise, the learned mapping remains a reasonable approximation but shows larger deviations from the oracle, particularly in the tails. The learned map is optimized to capture auxiliary statistics compatible with the GR2R objective. 

\pagebreak
In the correlated-noise setting, each coefficient of the $3\times 3$ kernel is treated as a free parameter within $h$, and the learned kernel approaches the overall structure of the true Gaussian blur kernel while not perfectly reproducing its symmetry. Such symmetry can, in principle, be imposed via a structured parameterization; however, empirical evidence indicates that this is not necessary to achieve satisfactory denoising performance. Overall, the same min--max objective adapts the recorruptor through its monotonic transport and learnable spatial kernel. The learned $h$ provides an interpretable proxy for the effective auxiliary noise statistics.

\subsection{Implicit noise-correlation tracking}
\label{sec:moment}

To assess the learned recorruption and the effect of the maximization in Eq.~\eqref{eq: lagrange minim}, an additional evaluation is conducted on the log-gamma denoising setting with $\ell=1.0$. During training, the following quantities are monitored:
\begin{align}
\mathcal{C}_{\varepsilon} &\triangleq \mathbb{E}_{\x,\w, \w'}\!\left[f(\y + \tau h(\w'))^\top g(\w) \right], \label{eq:c_eps}  \
\mathcal{C}_{h} \triangleq \mathbb{E}_{\x, \w,\w'}\!\left[\frac{1}{\tau} f(\y + \tau h(\w'))^\top h(\w')\right],  \\
\mathcal{C}_{\mathrm{\Delta}} &\triangleq \left| \mathbb{E}_{\x,\w,\w'}\!\left[f(\y+\tau h(\w'))^\top g(\w) - \frac{1}{\tau} f(\y + \tau h(\w') )^\top h(\w')\right]\right|. \label{eq:delta}
\end{align}
Here, $\mathcal{C}_{\varepsilon}$ is the \emph{noise-correlation term}, $\mathcal{C}_h$ is the \emph{auxiliary correlation term}, and $\mathcal{C}_{\mathrm{\Delta}}$ measures the \emph{residual gap} between both terms.

\begin{figure}[!htbp]
    \centering
    \includegraphics[width=0.78\linewidth,trim=0 0 538bp 0,clip]{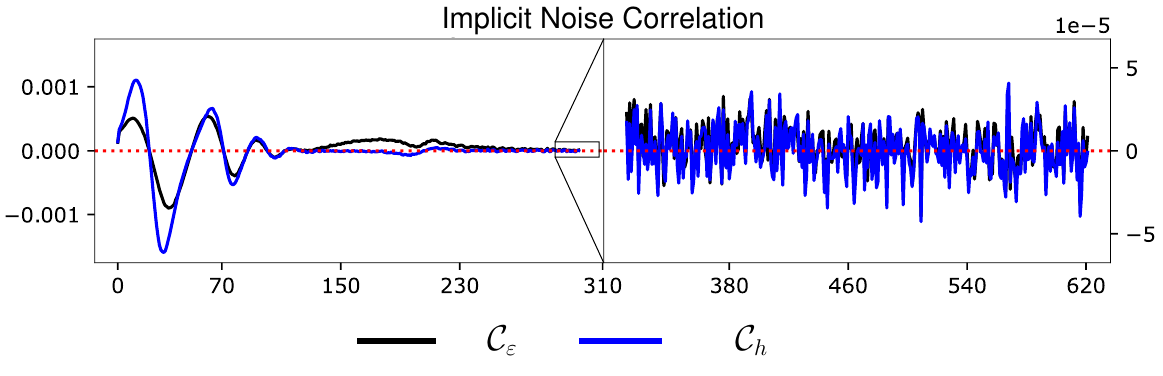}\\[0.4em]
    \includegraphics[width=0.78\linewidth,trim=538bp 0 0 0,clip]{imgs/correlation.pdf}
  \caption{
\textbf{Training dynamics of the implicit correlation terms.} \textit{Top:} noise correlation $\mathcal{C}_\varepsilon$ (\textbf{black}) and auxiliary correlation $\mathcal{C}_h$ ({\textbf{blue}}) versus epoch. \textit{Bottom:} residual gap $\mathcal{C}_{\mathrm{\Delta}}$ (\textbf{green}). Both inner-product terms settle near zero. The residual gap decreases and stabilizes near zero, supporting the empirical alignment described in Eq.~\eqref{eqn:equilibrium}.}
\label{fig:vanish}
\end{figure}

Figure~\ref{fig:vanish} reports these statistics across epochs. Two regimes are observed. Early in training, both $\mathcal{C}_{\varepsilon}$ and $\mathcal{C}_h$ exhibit oscillatory behavior as the min--max optimization evolves. The auxiliary correlation term (blue) settles near zero as training progresses, consistent with the intended effect of the saddle formulation, i.e., enforcing $\mathbb{E}\!\left[ \frac{1}{\tau} f(\y_1)^\top h(\w')\right] \approx 0$ without access to clean targets.

The noise-correlation term (black) also settles near zero and closely tracks $\mathcal{C}_h$. The residual gap $\mathcal{C}_{\mathrm{\Delta}}$ (green) decreases and stabilizes near zero, supporting the empirical alignment described in Eq.~\eqref{eqn:equilibrium}.

\end{document}